\documentclass[a4paper,12pt]{article}
\usepackage{amssymb}
\usepackage{amsfonts}
\usepackage{amsmath}
\usepackage{eurosym}
\usepackage{jheppub}
\usepackage{lineno}

\title{\boldmath Superspin Chains Solutions from 4D Chern-Simons Theory}

\author[1,2]{Y. Boujakhrout}
\author[1,2]{E.H Saidi}
\author[1,2]{R. Ahl Laamara}
\author[1,2]{L.B Drissi}

\affiliation[1]{LPHE-MS, Science Faculty, Mohammed V University in Rabat, Morocco}
\affiliation[2]{Centre of Physics and Mathematics, CPM- Morocco}
\emailAdd{boujakhroutyoussra@gmail.com}
\abstract{ As a generalisation of the correspondence linking 2D integrable systems with
4D Chern-Simons \textrm{(CS)} gauge theory, superspin chains are realized by
means of crossing electric and magnetic super line defects in the 4D CS with
super gauge symmetry. The oscillator realization of Lax operators solving
the RLL relations of integrability is obtained in the gauge theory by
extending the notion of Levi decomposition to Lie superalgebras. Based on
particular 3-gradings of Lie superalgebras, we obtain graded oscillator Lax
matrices for superspin chains with internal symmetries given by $A(m-1\mid
n-1)$, $B(m\mid n)$, $C(n)$ and $D(m\mid n)$.\newline
}
\input{tcilatex}
\begin{document}

\maketitle

\flushbottom

\section{Introduction}

\label{sec:intro}

A newly discovered shortcut towards the realization and study of integrable
systems is yielded by a four dimensional Chern-Simons gauge theory defined
on the product of a topological real plane $\Sigma $ and a holomorphic curve
$C,$ by the field action \cite{11}-\cite{1C}%
\begin{equation}
S_{4dCS}=\int_{\Sigma \times C}dz\wedge tr(\mathcal{A}\wedge d\mathcal{A}+%
\frac{2}{3}\mathcal{A}\wedge \mathcal{A}\wedge \mathcal{A})  \label{ac}
\end{equation}%
This field theory is characterized by a complexified gauge symmetry $G,$ and
a partial gauge connection with three bosonic components as $\mathcal{A}=dx%
\mathcal{A}_{x}+dy\mathcal{A}_{y}+d\bar{z}\mathcal{A}_{\bar{z}},$ valued in
the Lie algebra $g$ of the gauge symmetry$.$ Endowing this topological field
theory with crossing line defects allows to\ build two-dimensional solvable
models in $\Sigma ,$ and recover solutions and conserved quantities of these
lower dimensional models;\textrm{\ thus opening the door for interestings
findings enrishing the integrability literature \cite{1C1}-\cite{1H}}.%
\textrm{\ In particular, }the R-matrix describing the scattering of two
particles' worldlines \cite{p}-\cite{pp} is calculated from the 4D CS as the
crossing of two Wilson lines characterized by electrical charges given by
highest weights of $G$. In this image, each Wilson line is represented in
the topological plane by a line assimilated to the worldline of an
electrically charged particle whose internal quantum states are valued in
some representation $R$ of $g$ characterized by a highest weight $\lambda
_{R}$. Positions of these line defects in the complex $C$ correspond to
spectral parameters $z_{i}$ that play a major role in Yang-Baxter equation
and in the RTT realization of Yangian representations \cite{r},\cite{rr}.%
\newline
The integrable XXX spin chain \cite{s},\cite{ss} emerges in the 4D CS theory
defined on $\mathbb{R}^{2}\times C,$ as a set of parallel (vertical) Wilson
lines sitting on the chain nodes and carrying degrees of freedom of the
spins. The interaction between these spins is modelled by a horizontal 't
Hooft line perpendicularly crossing the Wilson lines \cite{costello}. The 't
Hooft line defect is a disorder operator \cite{t}-\cite{ttt} characterized
by a magnetic charge equivalent to a coweight $\mu $\ of $G;$ it acts like
an auxiliary oscillatory space such that its intersection with a Wilson line
at each node of the spin chain yields a Lax operator \cite{mar}. This
operator is a basic ingredient of the Bethe Ansatz approach \cite{q}-\cite%
{qq}; it operates on the quantum spaces and is a solution to the RLL
equation underlying the integrable spin chain. In the Gauge theory
formulation, the Lax operator is computed as the parallel transport of gauge
fields past the 't Hooft line. This key result was demonstrated in \cite%
{costello} for the particular case where the magnetic charge is given by a
minuscule coweight $\mu $ of $G.$ There, the authors linked the Levi
decomposition of the Lie algebra $g$\ to the dispersion of the gauge field
bundles above and under the 't Hooft line due\ the Dirac-like singularity
induced by the presence of this magnetic operator. The particularity of the
minuscule coweight is that it acts on the roots of $g$ with the eigenvalues $%
0,\pm 1$ \cite{minus} which decomposes the Lie algebra $g$ into three
subspaces as $n_{-1}\oplus l_{\mu }\oplus n_{+1}$. This is a Levi
decomposition of $g$ where the Levi subalgebra $l_{\mu }$\ has charge $0$
with respect to $\mu ,$ and the $n_{\pm 1}$\ are nilpotent subspaces given
by modules of $l_{\mu }$\ and carrying charges $\pm 1$ \cite{l},\cite{ll}.%
\newline
These algebraic features play a major role in this investigation because for
any 't Hooft line with minuscule magnetic charge $\mu $ of $G,$ the
corresponding Lax operator can be simply computed by the general formula $%
\mathcal{L}^{\mu }\left( z\right) =e^{X}z^{\mu }e^{Y},$ where $X=X_{i}b^{i}$
and $Y=Y^{i}c_{i}$ are elements of $n_{+}$ and $n_{-}$ \textrm{respectively}%
.\ The action of the coweight $\mu $ on the representation $R$ carried by
the Wilson line in question,\textrm{\ can be deduced by the branching rule
of }$R$\textrm{\ following from the Levi decomposition of }$g$.\ The
oscillator structure of the phase space of this L-operator follows from the
Levi decomposition properties;\textrm{\ the classical coordinates }$b^{i}\in
n_{+1}$\textrm{\ and }$c_{j}\in n_{-1}$\textrm{\ verify the Poisson bracket}$%
\  \left \{ b^{i},c_{j}\right \} _{PB}=\delta _{j}^{i}$\textrm{\ which yields
the commutator }at the quantum level \cite{costello},\cite{slmn}$.$ Based on
this interpretation, the minuscule Lax operators for the simply laced $A$
and $D$ type bosonic spin chains were first realized using the CS gauge
theory in \cite{costello}, and then in \cite{quiver,abcde}, in agreement
with solutions obtained using Yangian representations in \cite{FRASSEK1}-%
\cite{FRASSEK222}. Lax operators \textrm{for} bosonic spin chains with non
simply laced $B$ and $C$ type symmetries were recovered from 4D CS in \cite%
{abcde} and were compared with \cite{FRASSEK3}. The power of this 4D CS/
Integrability correspondence allowed also to build solutions for exceptional
spin chains, with internal symmetries described by the simply laced e$_{6}$
and e$_{7}$ algebras,\ which were lacking in the spin chain literature \cite%
{excep}. The missing exceptional e$_{8}$, \ f$_{4}$ and g$_{2}$ symmetries
in this bosonic \textrm{list }do not have minuscule coweights \cite{minus}.%
\newline
Regarding superspin chains with internal symmetry described by Lie
superalgebras, the generalization of the 4D CS/ Integrability correspondence
requires the equipment of a 4D Chern-Simons theory having super gauge
symmetry with super line defects carrying bosonic and fermionic degrees of
freedom \cite{nafiz}. This extension was motivated in \cite{slmn} by
uplifting from $SL(m)$\ to the $SL(m|n)$\ symmetry and by taking advantage
of the resemblance of their algebraic structure. The super Lax operators
characterizing the $sl(m|n)$\ superspin chain were calculated in the
framework of the $SL(m|n)$\ 4D CS by using a generalized formula similar to
the bosonic $L^{\mu }\left( z\right) =e^{X}z^{\mu }e^{Y}$. However, due to
the lack of the notion of minuscule coweight and Levi decomposition in the
superalgebras literature, a Dynkin diagram's node cutting method was used in
order to generate 3-gradings of the $sl(m|n)$\ Lie superalgebra. These
decompositions of the $sl(m|n)$\ family have similar properties to the Levi
decomposition such that the role of the minuscule coweight is played by the
cutted node. This approach allowed to construct explicit super L-operators
in terms of bosonic and fermionic oscillators of the phase space, in
agreement with the superspin chain literature solutions \cite{FRC} \textrm{%
obtained} using degenerate solutions of the graded Yang-Baxter equation \cite%
{kulish}.\newline
In this paper, we follow a similar approach to \cite{slmn} in order to build
oscillator realizations of super Lax operators for integrable superspin
chains classified by the basic $ABCD$\ Lie superalgebras. As for these Lie
superalgebras, one has several super Dynkin diagrams depending on the number
of fermionic roots and their ordering. Therefore, one distinguishes several
varieties of the ABCD superspin chains due to their link to the super
Dynkins \cite{embedding}. By considering a super Wilson line $W_{\xi _{z}}^{%
\boldsymbol{R}}$\ in a given super representation $\boldsymbol{R}$\ and a
super 't Hooft tH$_{\mathrm{\gamma }_{0}}^{\mathbf{\mu }}$\ with magnetic
charge $\mathbf{\mu ,}$ we calculate the super Lax operator describing their
crossing. For the $sl(m|n)$\ symmetry, we derive the super L-operators for
any coweight $\mathbf{\mu }$ of any super Dynkin diagram of the $\left(
m+n\right) !/m!n!$ possible graphs. We show that they agree with those
calculated by using the super Yangian representations \textrm{as a
verification of our approach}. For the $B(m|n)$, $C(n)$\ and $D(m|n)$\
\textrm{superspin chains, we give a family of solutions corresponding to
specific coweights of the distinguished super Dynkin diagrams leading to
Levi-like 3-gradings.}\newline
The presentation is as follows: In section $2$, we give basic tools of the
4D Chern-Simons theory with $SL(m|n)$ gauge symmetry, and the realization of
the $sl(m|n)$ superspin chain by means of super line defects. We describe
the super Lax operator construction for basic Lie superalgebras. In section $%
3$, we use this construction to build solutions for the RLL equations of the
$sl(m|n)$ superspin chain, and compare with known results of the literature.
Sections $4,5$ and $6$ are respectively dedicated to the building of super
Lax operators for superspin chains with $B(m|n)$\textrm{, }$C(n)$\ and $%
D(m|n)$\ symmetries. We end with a conclusion and discussions.\textrm{\ An
appendix is added to this version as a verification of the new
orthosymplectic solutions.}

\section{$sl(m|n)$ superspin chain in 4D CS}

\label{sec:2} In this section, we consider the standard A$_{\text{\textsc{bf}%
}}$-family of superspin chains based on the $sl(m|n)$ Lie superalgebra ($%
m\neq n$), to first introduce the basics of the present investigation, and
to complete partial results in literature with regards to the\textrm{\ }A$_{%
\text{\textsc{bf}}}$ class \textrm{\cite{FRC}}. The label \textsc{bf} refers
to chains with bosonic and fermionic degrees of freedom. This family of
integrable super systems is realized in the framework of the 4D Chern Simons
gauge theory having the $SL(m|n)$\ super gauge group. The superspin chain
families \textrm{B}$_{\text{\textsc{bf}}}$\textrm{, C}$_{\text{\textsc{bf}}}$%
\textrm{\ }and \textrm{D}$_{\text{\textsc{bf}}}$\ to be studied in the
forthcoming sections are realized in\ a similar fashion; and as such, the
algebraic basics of the line defects construction for all superchains are
only detailed for the case of the A$_{\text{\textsc{bf}}}$ super chain.%
\newline
To this end, it is interesting to recall that the A$_{\text{\textsc{bf}}}$
special family of superchains generalizes the well known family of $sl(n)$
spin chains termed below as the A$_{\text{\textsc{bose}}}$ family. The
generalised A$_{\text{\textsc{bf}}}$ has two basic features: First, it is
classified by the set of Lie superalgebras $g_{\text{\textsc{bf}}}$\ given
by the bi-integer series A$_{\text{\textsc{bf}}}^{m,n}\equiv sl(m|n)$
including $sl(n)$ and $sl(m)$ as bosonic subsectors. Second, for $sl(m|n)$
one distinguishes several types of superchains versus one ordinary $sl(n)$
chain in the bosonic case. This is due to the $\mathbb{Z}_{2}$-grading of
the \textsc{bf} gauge symmetry to be commented\textrm{\ on l}ater. For
example, given two positive integers $\left( m,n\right) $, one has
\begin{equation}
N_{n,m}^{\text{A}_{\text{\textsc{bf}}}}=\frac{\left( m+n\right) !}{m!n!}
\end{equation}%
varieties of $sl(m|n)$ superchains.\newline
To perform this study, we begin by briefly describing the \textrm{%
corresponding} four-dimensional gauge theory with $SL(m|n)$ local symmetry,
and its line defects that allow for the superspin chain realization. Then,
we introduce \textrm{a general formula for the computation of super Lax
operators directly from the superalgebra 3-gradings.}

\subsection{4D CS gauge theory with $SL(m|n)$ symmetry}

The field action describing 4D Chern-Simons theory with super $SL(m|n)$
gauge symmetry, living on the 4D manifold $M_{4}=\mathbb{R}^{2}\times C,$
with $\mathbb{R}^{2}$ the real plan parameterized by $(x,y)$ and $C$ an
holomorphic curve parameterized by $z$, is written in terms of the
supertrace of the CS 3-form \textrm{\cite{slmn}}%
\begin{equation}
\mathcal{S}_{{\small CS}}^{sl_{(m|n)}}=\int_{\mathbb{R}^{2}\times C}dz\wedge
str\left[ \mathcal{A}\wedge d\mathcal{A}+\frac{2}{3}\mathcal{A}\wedge
\mathcal{A}\wedge \mathcal{A}\right]  \label{7}
\end{equation}%
The 1-form gauge potential $\mathcal{A}$ is given by $\mathcal{A}_{x}dx+%
\mathcal{A}_{y}dy+\mathcal{A}_{\bar{z}}d\bar{z}$ where we have dropped the
component $\mathcal{A}_{z}dz$ because of the $dz$ factor in the\textrm{\
holomorphic volume form}. It is valued in the $sl(m|n)$ Lie superalgebra,
and \textrm{thus} expands like
\begin{equation}
\mathcal{A}=\sum_{\text{\textsc{ab}}}A^{\text{\textsc{ab}}}\mathcal{E}_{%
\text{\textsc{ab}}}  \label{gauge}
\end{equation}%
where $\mathcal{E}_{\text{\textsc{ab}}}$ are graded generators of $sl(m|n)$
obeying the graded commutation relations%
\begin{equation}
\left[ \mathcal{E}_{\text{\textsc{ab}}},\mathcal{E}_{\text{\textsc{cd}}%
}\right \} =\delta _{\text{\textsc{bc}}}\mathcal{E}_{\text{\textsc{ad}}%
}-\left( -\right) ^{\left \vert \mathcal{E}_{\text{\textsc{ab}}}\right \vert
\left \vert \mathcal{E}_{\text{\textsc{cd}}}\right \vert }\delta _{\text{%
\textsc{da}}}\mathcal{E}_{\text{\textsc{cb}}}  \label{gc}
\end{equation}%
with degree as
\begin{equation}
\left \vert \mathcal{E}_{\text{\textsc{ab}}}\right \vert \equiv \deg
\mathcal{E}_{\text{\textsc{ab}}}=\left \vert \text{\textsc{a}}\right \vert
+\left \vert \text{\textsc{b}}\right \vert ,\quad \func{mod}2
\end{equation}%
The supertrace of the Chern-Simons 3-form in (\ref{7}) is therefore written
in terms of the graded metric $g_{\text{\textsc{abcd}}}=str(\mathcal{E}_{%
\text{\textsc{ab}}}\mathcal{E}_{\text{\textsc{cd}}})$\ and the constant
structures $f_{\text{\textsc{abcdef}}}=str(\mathcal{E}_{\text{\textsc{ab}}}%
\mathcal{E}_{\text{\textsc{cd}}}\mathcal{E}_{\text{\textsc{ef}}})$ as follows%
\begin{equation}
str\left( \mathcal{A}\wedge d\mathcal{A}+\frac{2}{3}\mathcal{A}\wedge
\mathcal{A}\wedge \mathcal{A}\right) =g_{\text{\textsc{abcd}}}A^{\text{%
\textsc{ab}}}dA^{\text{\textsc{cd}}}+\frac{2}{3}f_{\text{\textsc{abcdef}}}A^{%
\text{\textsc{ab}}}A^{\text{\textsc{cd}}}A^{\text{\textsc{ef}}}
\end{equation}%
\textrm{The field equation of the gauge connection }$A$\textrm{\ in absence
of external charges is given by a vanishing 2-form field strength}%
\begin{equation*}
F=d\mathcal{A}+\mathcal{A}\wedge \mathcal{A}=0
\end{equation*}%
\textrm{In order to realize lower dimensional integrable super systems, we
need to introduce charges to the CS theory through super line defects, such
as the super Wilson }$W_{\xi _{z}}^{\boldsymbol{m|n}}$ \textrm{\cite%
{1A,slmn,2A}}. This topological defect is represented by a \textrm{line} $%
\xi _{z}$\ in $\mathbb{R}^{2},$ sitting in the position $z$ in $C$ along
which\textrm{\ }propagate quantum super states. The $W_{\xi _{z}}^{%
\boldsymbol{m|n}}$ can be imagined as an electrically charged line defect,
characterized by the fundamental representation $\boldsymbol{R=m|n}$\ of the
superalgebra $sl(m|n),$\ such that the electric charge is given by the
corresponding highest weight. The topological $W_{\xi _{z}}^{\boldsymbol{m|n}%
}$ is defined as the supertrace of the holonomy of the gauge field\textrm{\
along the line} $\xi _{z}$ like
\begin{equation}
W_{\xi _{z}}^{\boldsymbol{m|n}}=str_{\boldsymbol{m|n}}\left[ P\exp \left(
\oint_{{\xi _{z}}}\mathcal{A}\right) \right]
\end{equation}%
with $\mathcal{A}$ as in (\ref{gauge}).\newline
\textrm{Besides} the Wilson $W_{\xi _{z}}^{\boldsymbol{m|n}},$\textrm{\ we
can also introduce }a magnetically charged super 't Hooft line \cite{t}-\cite%
{ttt}, denoted here as tH$_{\mathrm{\gamma }_{z^{\prime }}}^{\mathbf{\mu }}$%
. This is also a topological line defect, which is implemented\textrm{\ in
the 4D CS as a line} $\mathrm{\gamma }_{z^{\prime }}$\ extending in the
space $\mathbb{R}^{2},$\ and living at a point $z^{\prime }$ in $C.$ The tH$%
_{\mathrm{\gamma }_{z^{\prime }}}^{\mathbf{\mu }}$ is a disorder operator
that carries a magnetic charge\ given by a coweight $\mathbf{\mu }$ of the $%
SL(m|n)$ supergroup; and is identified with the parallel transport of gauge
field bundles past the line. By taking $\mathrm{\gamma }_{z^{\prime }}$ as
the x-axis in $\mathbb{R}^{2}$ and $z^{\prime }=0$, we can write the 't
Hooft line observable as%
\begin{equation}
\mathcal{L}^{\mathbf{\mu }}(z)=P\exp \left( \int_{y}\mathcal{A}_{y}(z)\right)
\label{L1}
\end{equation}%
where the transport of the gauge fields is measured from $y<0$ to $y>0.$
Actually, just like in the bosonic case \cite{costello}\textrm{, the
magnetically charged super 't Hooft line defect is classically defined\ such
that its presence in a 4D CS with super gauge symmetry }$G_{\text{\textsc{bf}%
}}$\textrm{, deforms the field action like}%
\begin{equation}
\mathcal{S}_{{\small CS}}^{G_{\text{\textsc{bf}}}}+\mathcal{S}_{{\small int}}%
\left[ \text{tH}_{\mathrm{\gamma }_{z^{\prime }}}^{\mathbf{\mu }}\right]
\end{equation}%
\textrm{Here, the field strength }$\mathcal{F}$\textrm{\ is no longer flat }$%
(\mathcal{F}\neq 0),$\textrm{\ and should look like a Dirac monopole with
non trivial first Chern class}%
\begin{equation}
c_{1}=\int_{\mathfrak{C}}\mathcal{F}
\end{equation}%
\textrm{where the surrounding of the line is conveniently viewed as a
cylinder }$\mathfrak{C}$\textrm{\ instead of a sphere, due to the mixed
nature of the theory \cite{costello}, and the coweight }$\mathbf{\mu }%
:U(1)\rightarrow G_{\text{\textsc{bf}}}$\textrm{\ serves to map the }$U(1)$%
\textrm{\ Dirac monopole into a non-abelian }$G_{\text{\textsc{bf}}}$\textrm{%
\ Dirac monopole. In this case, the gauge field defines a }$G_{\text{\textsc{%
bf}}}-$ \textrm{bundle on the surrounding of the 't Hooft line which is also
defined as the mapping of the abelian }$U(1)$\textrm{\ bundle by the action
of the magnetic graded coweight }$\mathbf{\mu }.$\textrm{\ This gauge
configuration is described on three regions that can be identified with the
bases and the contour of a cylinder }$\mathfrak{C}$\textrm{\ in }$(x,y,z)$%
\textrm{. Above and under the} tH$_{\mathrm{\gamma }_{0}}^{\mathbf{\mu }}$%
\textrm{, we have trivialized bundles in the regions }$U_{I}=\left \{ y\leq
0,z\sim 0\right \} $\textrm{\ and }$U_{II}=\left \{ y\geq 0,z\sim 0\right \}
,$ \textrm{which are locally glued by a transition function (isomorphism)
near the intersection }$U_{I}\cap U_{II}=\left \{ y=0,z\sim 0\right \}
\simeq \mathrm{\gamma }_{0}.$\  \textrm{The\ trivial bundles in }$U_{I,II}$%
\textrm{\ should be holomorphic }$G_{\text{\textsc{bf}}}-$\textrm{\ valued
functions that are regular at }$z=0;$\textrm{\ they are given by gauge
transformations }$\mathfrak{g}_{I}(z),$ $\mathfrak{g}_{II}(z)$\textrm{\ with
monopole charges}$.$\textrm{\ The transition function serves as a parallel
transport from }$U_{I}$\textrm{\ to }$U_{II},$\textrm{\ and is locally equal
to the Dirac singularity }$z^{\mathbf{\mu }}$ \cite{arXiv:0812.0221}\textrm{%
. The L-operator (\ref{L1})\ measuring this gauge behaviour near }tH$_{%
\mathrm{\gamma }_{0}}^{\mathbf{\mu }}$\textrm{\ is therefore given by the
formula}%
\begin{equation}
\mathcal{L}^{\mathbf{\mu }}(z)=\mathfrak{g}_{I}(z)z^{\mathbf{\mu }}\mathfrak{%
g}_{II}(z)  \label{L2}
\end{equation}%
Notice that the tH$_{\mathrm{\gamma }_{0}}^{\mathbf{\mu }}$ is in fact
sitting at the end of a Dirac string linking it to another 't Hooft line
living at $z=\infty ,y=0$ and having the opposite magnetic charge $-\mathbf{%
\mu }$. \textrm{The gauge behaviour near }$z=\infty $\textrm{\ is
equivalently evaluated as}%
\begin{equation}
\mathfrak{g}_{III}(z^{-1})z^{-\mathbf{\mu }}\mathfrak{g}_{IV}(z^{-1})
\end{equation}%
$\mathfrak{g}_{III}(z^{-1})$\textrm{\ and }$\mathfrak{g}_{IV}(z^{-1})$%
\textrm{\ are }$G_{\text{\textsc{bf}}}-$\textrm{\ valued holomorphic
functions that need to be regular and equal to the identity at }$z=\infty $
\cite{costello}$.$\textrm{\ However, this detail will be omitted in the
present inquiry, and we will be considering a 't Hooft line }tH$_{\mathrm{%
\gamma }_{0}}^{\mathbf{\mu }}$\textrm{\ which is to be understood as coupled
to a }tH$_{\mathrm{\gamma }_{\infty }}^{\mathbf{-\mu }}.$\textrm{\ The
parallel transport of the gauge field in the presence of this double line
having charges at }$z=0$\textrm{\ and }$z=\infty $\textrm{\ is of the form}%
\begin{equation}
\mathcal{L}^{\mathbf{\mu }}(z)=A(z)z^{\mathbf{\mu }}B(z)  \label{formula}
\end{equation}%
\textrm{where }$A(z)$\textrm{\ and }$B(z)$\textrm{\ are }$G_{\text{\textsc{bf%
}}}-$\textrm{\ valued functions\ verifying the appropriate singularity
constraints at zero and infinity.}\newline
\textrm{In analogy to the parallel transport of a 't Hooft line in a bosonic
Chern-Simons theory \cite{costello}, we should be able to explicitly realize
the functions of the formula (\ref{formula}) in terms of a 3-grading of the
Lie superalgebra }$g_{\text{\textsc{bf}}}$\textrm{. This was shown for the }$%
SL(m|n)$\textrm{\ symmetry in \cite{slmn} (see Appendix A)\ and will be
generalized in the next section\ for any superalgebra }$g_{\text{\textsc{bf}}%
}.$

\subsection{Oscillator realization in 4D CS theory}

\begin{figure}[h]
\begin{center}
\includegraphics[width=13cm]{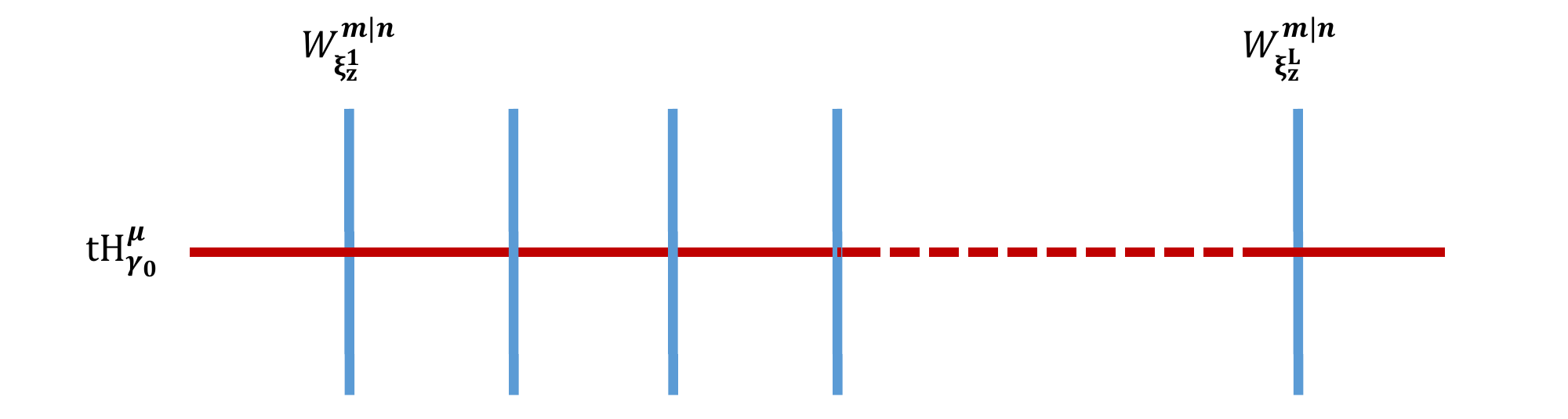}
\end{center}
\caption{Realization of an $sl(m|n)$ superspin chain of $L$ nodes in the
fundamental representation using super line defects in the 4D Chern Simons.}
\label{chain}
\end{figure}

\begin{figure}[h]
\begin{center}
\includegraphics[width=11cm]{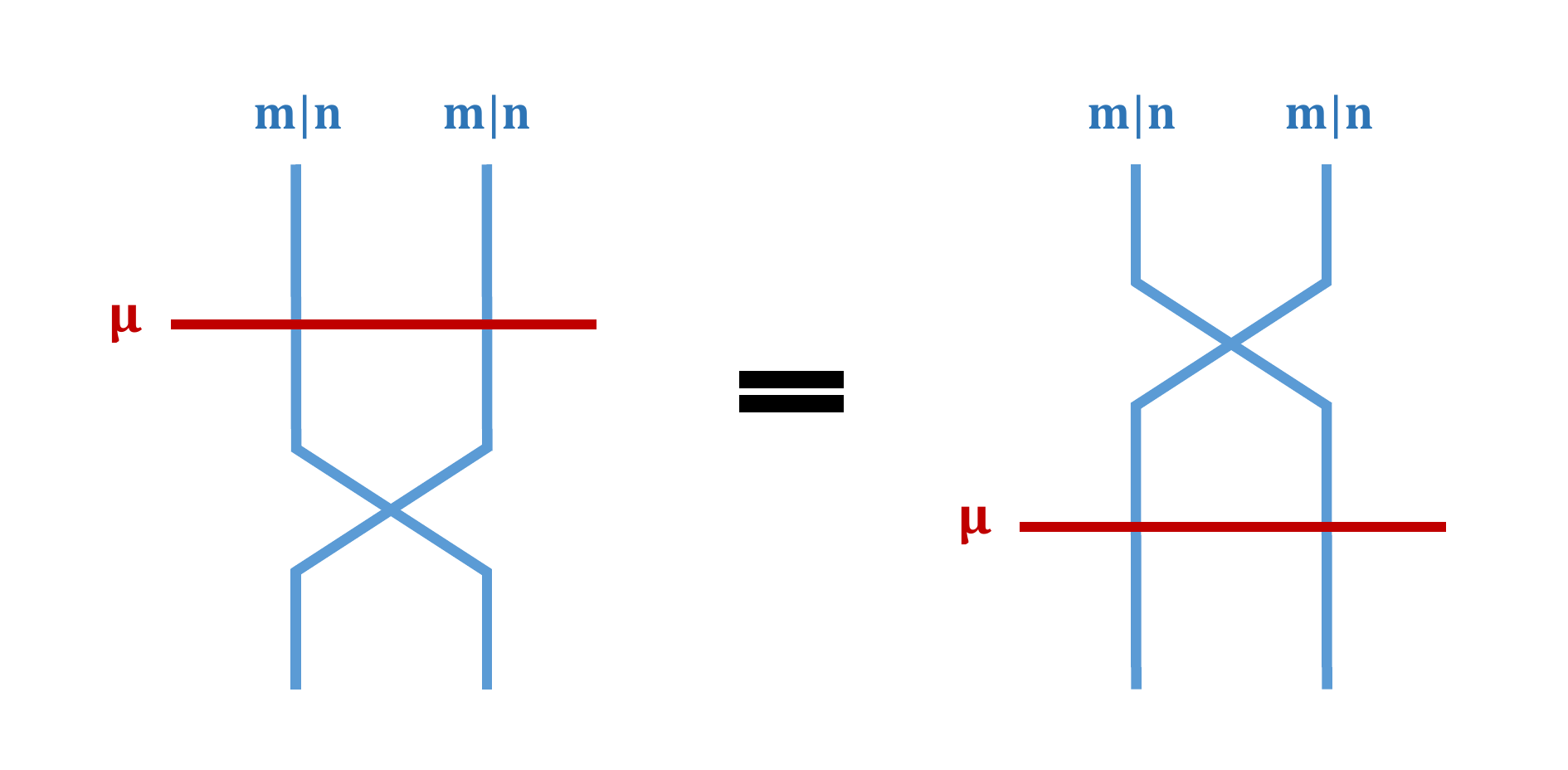}
\end{center}
\caption{Graphic representation of the RLL equation in terms of intersecting
line defects in $SL(m|n)$ 4D CS theory.}
\label{RTT}
\end{figure}

We \textrm{focus now on linking the graded gauge theory and its ingredients}
presented above to the integrable superchain systems we are concerned about
here. We describe the super 4D CS/ superchain correspondence by focussing on
the $SL(m|n)$\ symmetry, in order to introduce the super Lax operator as a
solution to the RLL equation of \textrm{integrability, and its }%
interpretation and computation in the gauge theory.\newline
To begin, the super 4D CS/ superchain correspondence we are considering here
is an extension of the well known bosonic 4D CS/ Integrability \textrm{%
linking a spin chain with bosonic internal symmetry }$sl(n)$\textrm{\ to 4D
Chern-Simons theory with gauge group }$SL(n).$ This bosonic correspondence
is also valid for other gauge symmetries $G_{\text{\textsc{bose}}}$ given by
Cartan classification of finite dimensional Lie algebras including $A_{n},$ $%
B_{n},$ $C_{n},$ $D_{n}$ and the exceptional ones. \textrm{Alongside} this
interesting result, it \textrm{was} also \textrm{conjectured} that the
bosonic \textrm{correspondence} extends to superspin chains where several
checks \textrm{were successfully carried out} \cite{nafiz},\cite{slmn}.%
\newline
In this Fermi/Bose generalisation, superchains are characterised by
superspin representations of Lie superalgebras $g_{\text{\textsc{bf}}}$
splitting like
\begin{equation}
g_{\text{\textsc{bf}}}=g_{\bar{0}}\oplus g_{\bar{1}}
\end{equation}%
with $g_{\bar{0}}$ a bosonic Lie algebra and $g_{\bar{1}}$ a module of it.
As such, basic integrable superspin chains fall into families classified by
the basic \textrm{complexified} Lie superalgebras $g_{\text{\textsc{bf}}}$\
as listed below \textrm{\cite{dictionnary},\cite{str}}%
\begin{equation}
\begin{tabular}{c|cc}
$g_{\text{\textsc{bf}}}$ & even part $g_{\bar{0}}$ & odd part $g_{\bar{1}}$
\\ \hline \hline
$A(m|n)$ & $A_{m}\oplus A_{n}\oplus gl(1)$ & $(\bar{m},n)\oplus (m,\bar{n})$
\\ \hline
$B(m|n)$ & $B_{m}\oplus C_{n}$ & $(2m+1,2n)$ \\ \hline
$C(n)$ & $C_{n-1}\oplus gl(1)$ & $(2n-2)\oplus (2n-2)$ \\ \hline
$D(m|n)$ & $D_{m}\oplus C_{n}$ & $(2m,2n)$ \\ \hline \hline
\end{tabular}%
\end{equation}%
In general, an integrable superspin chain with intrinsic \textrm{symmetry}
given by\textrm{\ the} superlagebra $g_{\text{\textsc{bf}}}$ is linked to
the 4D Chern-Simons theory \textrm{characterized by} the corresponding super
gauge group $G_{\text{\textsc{bf}}},$ \textrm{and }a field action $\mathcal{S%
}_{{\small CS}}^{g_{\text{\textsc{bf}}}}[\mathcal{A}]$ as eq(\ref{7})
\textrm{where the} potential $\mathcal{A}=\mathcal{A}_{x}dx+\mathcal{A}%
_{y}dy+\mathcal{A}_{\bar{z}}d\bar{z}$ \textrm{expands in terms of the
generators of }$g_{\text{\textsc{bf}}}$ as in eq(\ref{gauge}). The extension
of the bosonic gauge/ Integrability \textrm{correspondence} to graded
symmetries is permitted by the implementation of the super line defects
introduced previously. In the $SL(m|n)$ generalisation, the super Wilson
line $W_{\xi _{z}}^{\boldsymbol{m|n}}$\ carries degrees of freedom that
allow to describe the quantum states of a \textrm{super atom with superspin
valued in the representation }$\boldsymbol{m|n}$ \cite{embedding}\textrm{,
i.e.} with $sl(m|n)$ internal symmetry.\textrm{\ Therefore, the} integrable $%
sl(m|n)$\textrm{\ }superchain with $\mathcal{N}$\ super atoms \textrm{can be}
realised in the 4D CS theory by placing $\left( i\right) $ $L$ vertical
Wilson lines $W_{\xi _{z}^{i}}^{\boldsymbol{m|n}}$\ at \textrm{every} node $%
\nu _{i}$ ($1\leq i\leq L$) of the superspin chain\ such that
\begin{equation}
\xi _{z}^{i}=\left( z_{i},x_{i},\mathbb{R}\right) \text{ \ }\qquad \text{%
with }\qquad \left \{
\begin{array}{c}
z_{i}=z \\
x_{i}<x_{i+1} \\
-\infty <y<\infty%
\end{array}%
\right.
\end{equation}%
\ and $\left( ii\right) $ an horizontal 't Hooft line tH$_{\mathrm{\gamma }%
_{0}}^{\mathbf{\mu }}$ with $\mathrm{\gamma }_{0}=\left( z_{0},\mathbb{R}%
,y_{0}\right) $ that can be thought of as filling the x-axis ($y_{0}=0$) in $%
\mathbb{R}^{2}$ and sitting at $z_{0}=0$ in $\mathbb{CP}^{1}$. These super
line defects intersect in the topological plane $\mathbb{R}^{2}\left(
x,y\right) $ as depicted in \textbf{Figure \ref{chain}},
where the 't Hooft line plays\ the role of a transfer matrix modeling the
interactions between the electrically charged super atoms\ along the chain.
Following this lattice system realization, each intersection of an electric $%
W_{\xi _{z}}^{\boldsymbol{m|n}}$ \textrm{carrying the vector quantum space} $%
\boldsymbol{m|n}$\ with the magnetic tH$_{\mathrm{\gamma }_{0}}^{\mathbf{\mu
}}$ \textrm{carrying an auxiliary space}, yields the super Lax operator for
the corresponding node of the superspin chain. This coupling operator acts
on the tensor product of $End(\boldsymbol{m|n})$ of the Wilson and the
algebra {\large A} of functions in the phase space of the 't Hooft line%
\textrm{. It is nothing but the L-operator (\ref{L1}})\textrm{\ which
describes the paralell transport of gauge fields (given here by the fields
in }$\boldsymbol{m|n}$\textrm{, travelling along the Wilson line) past the
't Hooft line carrying the magnetic charge }$\mathbf{\mu }$\textrm{\ that
acts on the space }$\boldsymbol{m|n}$; it can be simply labeled by the
representation $\mathbf{R}=\boldsymbol{m|n}$ and the coweight $\mathbf{\mu .}
$\newline
The quantum integrability of this system is encoded in the RLL equation
verified by the L-operator with matrix realisation $L_{n}^{m}\left( z\right)
$ obeying,\textrm{\ }%
\begin{equation}
R_{rs}^{ik}\left( z-w\right) L_{j}^{r}\left( z\right) L_{l}^{s}\left(
w\right) =L_{r}^{i}\left( w\right) L_{s}^{k}\left( z\right)
R_{jl}^{rs}\left( z-w\right)  \label{RLLL}
\end{equation}%
where $R_{rs}^{ik}\left( z-w\right) $ is the usual R-matrix of the
Yang-Baxter equation. \textrm{Notice here that the Wilson/ 't Hooft crossing
was shown to verify this equivalence by field theory analysis \cite%
{costello,1B}. The }RLL equation has a remarkable graphical representation
given by \textbf{Figure \ref{RTT}. }\textrm{It is diagrammatically verified
thanks to the diffeomorphism invariance of the theory in 4D, where the 't
Hooft line can be freely moved past the Wilsons' crossing since it sits at a
different position }$z$\textrm{\ in }$C$.
\textrm{Now, in order to explicitly compute the super\ Lax operators of a
superspin chain in the framework of the 4D CS theory, we follow the approach
of \cite{costello} which builds the oscillator realization for a Lax
operator of a spin chain with internal symmetry }$g$\textrm{\ by considering
the magnetic charge of the 't Hooft line as a minuscule coweight \cite{minus}%
. The minuscule coweight }$\mu $\textrm{\ acts on the algebra elements with
eigenvalues }$0,\pm 1$\textrm{\ and induces a Levi decomposition of }$g$%
\textrm{\ as}%
\begin{equation}
g=n_{-}\oplus l_{\mu }\oplus n_{+}  \label{olo}
\end{equation}%
\textrm{where }$l_{\mu }$\textrm{\ is the Levi subalgebra containing
elements of charge 0 with respect to }$\mu ,$\textrm{\ and} $n_{\pm }$
\textrm{are nilptent subspaces carrying charges }$\pm 1.$\textrm{\ These
properties are described by the Levi constraints}%
\begin{equation}
\left[ \mu ,n_{\pm }\right] =\pm n_{\pm };\qquad \left[ n_{+},n_{-}\right] =%
\mathbb{C}\mu ;\qquad \left[ n_{\pm },n_{\pm }\right] =0  \label{kk}
\end{equation}%
\textrm{In this case, the splitting of gauge bundles in the L-operator
formula as in eq(\ref{formula}) is identified on the Lie algebra level with
the Levi decomposition with respect to the magnetic coweight }$\mu $\textrm{%
. The Lax operator }$\mathcal{L}^{\mathbf{\mu }}(z)$\textrm{\ solving the
RLL equation for a chain of spins in the representation} $R,$\textrm{\
associated to the node corresponding to the coweight }$\mu $\textrm{\ in the
Dynkin diagram can be directly computed using the expression}%
\begin{equation}
\mathcal{L}_{R}^{\mu }=e^{X}z^{\mu }e^{Y}
\end{equation}%
\textrm{Here, }$X=X_{i}b^{i}\in n_{+}$ and $Y=Y^{i}c_{i}\in n_{-}.$\textrm{\
The operator }$\mu $\textrm{\ acts on the representation }$R$\textrm{\ by
dividing it into subspaces carrying charges with respect to }$\mu $\textrm{\
such that the trace on the total space }$R$\textrm{\ is zero, }$%
tr_{R}\mu =0.$\textrm{\ These charges can also be deduced from the branching
rules \cite{slansky} of }$R$\textrm{\ following from the Levi decomposition (%
\ref{olo}).}\newline
\textrm{Now in order to compute the super Lax operators corresponding to
integrable superspin chains, one may be tempted to just generalise the
construction done for the integrable bosonic chain; but this poses a problem
because the notion of Levi decomposition and minuscule coweight are not yet
known for Lie superalgebras.\ We propose here to circumvent this difficulty
by using superalgebra 3-gradings generated by the method of (extended)
Dynkin diagram \cite{dynkin}. This method consists of deducing the possible
decompositions of a superalgebra by cutting nodes from a corresponding super
Dynkin diagram. These decompositions are\ motivated by the bosonic case
where the Levi decompositions of the }$ABCDE$\textrm{\ Lie algebras can also
be read on the Dynkin diagram level as the cutting of specific nodes
associated to minuscule coweights \cite{abcde}, \cite{excep}, \cite{quiver}.
The superalgebras decompositions we are interested in here are 3-gradings
having similar properties to Levi decompositions, which allows for the
oscillator realization of the auxiliary phase space. This type of 3-gradings
act as Levi decompositions for superalgebras, and were used in \cite{slmn}\
to recover super Lax operators for the }$sl(m|n)$\textrm{\ superchain from
4D CS, which were verified to agree with the literature.}\newline
The Dynkin diagram cutting method is described in the series of papers \cite%
{van}-\cite{van2} where the results are listed for all nodes of the Dynkin
diagrams of basic superalgebras. \textrm{Given }a super Dynkin diagram of a
superalgebra $\mathbf{g}_{\text{\textsc{bf}}}$, one can determine all
regular subalgebras $\mathbf{g}_{0}$\ of $\mathbf{g}_{\text{\textsc{bf}}}$\
by consecutive nodes cutting from the Dynkin (and extended) diagrams. This
yields decompositions of the superalgebra $\mathbf{g}_{\text{\textsc{bf}}}$\
having a general 5-grading form%
\begin{equation}
\mathbf{g}_{\text{\textsc{bf}}}=\mathbf{g}_{-2}\oplus \mathbf{g}_{-1}\oplus
\mathbf{g}_{0}\oplus \mathbf{g}_{+1}\oplus \mathbf{g}_{+2}
\end{equation}%
where $\mathbf{g}_{\pm k}$\ with $k=1,2$\ are\ $g_{0}$-modules determined by
means of representation techniques and where\textrm{\ }%
\begin{equation}
w(\mathbf{g}_{+k})=\mathbf{g}_{-k}
\end{equation}%
with $w$\ being the standard antilinear anti-involutive mapping of the Lie
superalgebra $g_{\text{\textsc{bf}}}$. For the purposes of our study, we are
only interested in the case where $\mathbf{g}_{\pm 2}=0$, i.e. in
decompositions of Lie superalgebras\ that look like%
\begin{equation}
\mathbf{g}_{\text{\textsc{bf}}}=\mathbf{g}_{-1}\oplus \mathbf{g}_{0}\oplus
\mathbf{g}_{+1}  \label{form}
\end{equation}%
These 3-gradings result from the cutting of specific nodes in the Dynkin
diagram, and are analogous to Levi decompositions for bosonic Lie algebras
because we have \cite{van}-\cite{van2}%
\begin{equation}
\left[ \mathbf{g}_{j},\mathbf{g}_{l}\right] =\mathbf{g}_{j+l}
\end{equation}%
\textrm{\ }with $j,l=0,\pm 1.$\ In what follows, we will interpret these
values $0,\pm 1$\textrm{\ associated the subspaces of }$\mathbf{g}_{\text{%
\textsc{bf}}},$\ as Levi charges with respect to the\ coweight $\mathbf{\mu }
$ corresponding to the cut node and acting as a "minuscule" coweight\textrm{%
\ of }$\mathbf{g}_{\text{\textsc{bf}}}$.\newline
\textrm{Now having a Levi-like decomposition of the superalgebra }$\mathbf{g}%
_{\text{\textsc{bf}}}$\textrm{\ of the form (\ref{form}) that we write as,}%
\begin{equation}
\mathbf{g}_{\text{\textsc{bf}}}=\mathbf{N}_{+}\oplus \boldsymbol{l}_{\mathbf{%
\mu }}\oplus \mathbf{N}_{-}
\end{equation}%
\textrm{which is associated to a minuscule-like coweight }$\mathbf{\mu }$
\textrm{with,}%
\begin{equation}
\left[ \mathbf{\mu },\mathbf{N}_{\pm }\right] =\pm \mathbf{N}_{\pm };\qquad %
\left[ \mathbf{N}_{+},\mathbf{N}_{-}\right \} =\mathbb{C}\mathbf{\mu }%
;\qquad \left[ \mathbf{N}_{\pm },\mathbf{N}_{\pm }\right \} =0
\end{equation}%
\textrm{we can show that the super Lax operator formula can be factorized
and directly calculated from elements of this decomposition, just like the
bosonic construction. In this case, an element of the gauge group }$G_{\text{%
\textsc{bf}}}$\textrm{\ is written as}%
\begin{equation}
e^{\mathbf{g}_{\text{\textsc{bf}}}}=e^{\mathbf{N}_{+}}e^{\boldsymbol{l}_{%
\mathbf{\mu }}}e^{\mathbf{N}_{-}}  \label{hjh}
\end{equation}%
\textrm{Notice here that elements of the graded nilpotents }$\mathbf{N}_{+}$%
\textrm{\ and }$\mathbf{N}_{-}$\textrm{\ expand in terms of bosonic as well
as fermionic generators, linked respectively to bosonic, and fermionic
coordinates that form graded oscillators; as we will see later on.}\newline
\textrm{By taking (\ref{hjh})\ into account, the functions }$A(z)\in G_{%
\text{\textsc{bf}}}$\textrm{\ and }$B(z)\in G_{\text{\textsc{bf}}}$\textrm{\
appearing in the formula (\ref{formula}) should be factorized as follows}%
\begin{equation}
\begin{tabular}{lll}
$A(z)=e^{a_{+}(z)}A^{0}(z)e^{a_{-}(z)}$ & ,\qquad & $A^{0}(z)=e^{a_{0}(z)}$
\\
$B(z)=e^{b_{+}(z)}B^{0}(z)e^{b_{-}(z)}$ & ,\qquad & $B^{0}(z)=e^{b_{0}(z)}$%
\end{tabular}%
\end{equation}%
\textrm{with}%
\begin{equation}
\begin{tabular}{l}
$a_{+}(z),b_{+}(z)\in \mathbf{N}_{+}$ \\
$a_{-}(z),b_{-}(z)\in \mathbf{N}_{-}$ \\
$a_{0}(z),b_{0}(z)\in \boldsymbol{l}_{\mathbf{\mu }}$%
\end{tabular}%
\end{equation}%
\textrm{By imposing the singularity conditions at }$z=0$\textrm{\ and }$%
z=\infty $\textrm{\ for these functions, we can bring the expression (\ref%
{formula})\ into the form}%
\begin{equation}
\mathcal{L}_{\mathbf{R}}^{\mathbf{\mu }}=e^{X}z^{\mathbf{\mu }}e^{Y},\qquad
X\in \mathbf{N}_{+},\qquad Y\in \mathbf{N}_{-}
\end{equation}%
\textrm{where\ }$X$\textrm{\ is valued in }$\mathbf{N}_{+},$\textrm{\ and }$%
Y $\textrm{\ belongs to }$\mathbf{N}_{-}.$\textrm{\ This calculation is
detailed for the }$sl(m|n)$\textrm{\ symmetry in Appendix A of \cite{slmn};
it equivalently holds for any gauge super group }$G_{\text{\textsc{bf}}}$%
\textrm{.\ In general, the }super Lax operator construction for any Lie
superalgebra $\mathbf{g}_{\text{\textsc{bf}}}$\ is described through the
following steps :

\textbf{A)} \emph{3-grading of the superalgebra }$\mathbf{g}_{\text{\textsc{%
bf}}}$\textrm{\newline
}We consider the Lie superalgebra $\mathbf{g}_{\text{\textsc{bf}}}$\ with a
3-grading%
\begin{equation}
\mathbf{g}_{\text{\textsc{bf}}}=\mathbf{N}_{+}\oplus \boldsymbol{l}_{\mathbf{%
\mu }}\oplus \mathbf{N}_{-}  \label{dec}
\end{equation}%
\textrm{obtained by a node cutting from a Dynkin diagram such that }$\mathbf{%
\mu }$\textrm{\ is the coweight of }$\mathbf{g}_{\text{\textsc{bf}}}$\textrm{%
\ associated to the deleted node. This coweight acts as a "super minuscule"
coweight and the three algebraic blocks in (\ref{dec})\ are as described
below:}

\begin{itemize}
\item \textrm{The }$\boldsymbol{l}_{\mathbf{\mu }}$\textrm{\ is a regular
Lie sub-(super)algebra of }$g_{\text{\textsc{bf}}}$\textrm{\ with elements
carrying charge 0 with respect to the coweight }$\mathbf{\mu }$\textrm{; it
plays the role of a Levi subalgebra.}
\begin{equation}
\left[ \mathbf{\mu ,}\boldsymbol{l}_{\mathbf{\mu }}\right] =0
\end{equation}%
In fact, the $\boldsymbol{l}_{\mathbf{\mu }}$ is always given by a direct
sum
\begin{equation}
\boldsymbol{l}_{\mathbf{\mu }}=\mathfrak{l}\mathbf{_{\mathbf{\mu }}}\oplus
\mathbb{C}\mathbf{\mu }
\end{equation}%
where $\mathbb{C}\mathbf{\mu }$ is associated to the cutted node\textbf{.}\
This can be visualized in the example of the bosonic\textrm{\ }$sl(p),$
where the cutting of the last node\textrm{\ }$\widetilde{\alpha }_{p-1}$\
from $\mathfrak{D}[sl(p)]$\textrm{\ }having\textrm{\ }$\left( p-1\right) $%
\textrm{\ }nodes \textrm{yields} two pieces:\textrm{\ }$\left( i\right) $\
the $\mathfrak{D}[sl(p-1)]$ of the Lie algebra $sl(p-1)$ having $p-2$ nodes
thought of as $l_{\mathbf{\mu }_{p-1}}$, and $\left( ii\right) $\ an
isolated node \textrm{\{}$\widetilde{\alpha }_{p-1}$\textrm{\}}
corresponding to\textrm{\ }$\mathbb{C}\mu _{p-1}$ due to $<\mu _{p-1},%
\widetilde{\alpha }_{p-1}>=1$\textrm{. }This $\mathbb{C}\mu _{p-1}$\ is
given by the abelian $gl\left( 1\right) $\ in the resulting Levi
decomposition reading as $sl(p)=n_{+}\oplus l_{\mu _{p-1}}\oplus n_{-}$%
\textrm{\ }with\textrm{\ }$l_{\mu _{p-1}}=sl(p-1)\oplus gl\left( 1\right) ,$
$\mathfrak{l}_{\mu _{p-1}}=sl(p-1)$\ and $n_{\pm }=p-1.$

\item The remaining elements of the decomposition (\ref{dec}) (i.e: elements
in $\mathbf{N}=\mathbf{g}_{\text{\textsc{bf}}}\backslash \boldsymbol{l}_{%
\mathbf{\mu }}$) are nilpotent; they are given by $\boldsymbol{l}_{\mathbf{%
\mu }}$-modules that carry charges $\pm 1$ with respect to the $\mathbf{\mu }
$.%
\begin{equation}
\left[ \mathbf{\mu ,N}_{\pm }\right] =\pm \mathbf{N}_{\pm }  \label{cond}
\end{equation}%
The two graded subspaces $\mathbf{N}_{\pm }$ mutually supercommute%
\begin{equation}
\left[ \mathbf{N}_{+},\mathbf{N}_{+}\right \} =0\qquad ,\qquad \left[
\mathbf{N}_{-},\mathbf{N}_{-}\right \} =0
\end{equation}%
They moreover verify the generalized Levi-like constraint%
\begin{equation}
\left[ \mathbf{N}_{+}\mathbf{,N}_{-}\right \} \subset \boldsymbol{l}_{%
\mathbf{\mu }}  \label{darb}
\end{equation}
\end{itemize}

\textbf{B)} \emph{Branching of representations of }$\mathbf{g}_{\text{%
\textsc{bf}}}$\newline
Under the decomposition (\ref{dec}), a representation $\mathbf{R}$ of the $%
\mathbf{g}_{\text{\textsc{bf}}}$ splits into a direct sum of irreducible
representations $\mathfrak{R}_{q_{k}}$ of the superalgebras in $\boldsymbol{l%
}_{\mathbf{\mu }}=\mathfrak{l}\mathbf{_{\mathbf{\mu }}}\oplus \mathbb{C}%
\mathbf{\mu }.$ These subspaces $\mathfrak{R}_{q_{k}}$ \textrm{carry charges
}$q_{k}$\textrm{\ with respect to }$\mu ,$\textrm{\ that can be identified
in the bosonic case from known branching rules. In general, we write}%
\begin{equation}
\mathbf{R}=\sum \limits_{k}\mathfrak{R}_{q_{k}}\qquad ,\qquad \left[ \mathbf{%
\mu ,}\mathfrak{R}_{q_{k}}\right] =q_{k}\mathfrak{R}_{qk}
\end{equation}%
The action of the coweight $\mathbf{\mu }$ on the representation $\mathbf{R}$
can be therefore defined as follows%
\begin{equation}
\mathbf{\mu }=\sum_{k}q_{k}\Pi _{k}\qquad ,\qquad \sum_{k}\Pi _{k}=I_{id}
\label{cow}
\end{equation}%
where $\Pi _{k}$\textrm{\ is the projector on the subspace} $\mathfrak{R}%
_{q_{k}}$ and $q_{k}$ is often termed as the Levi-charge.\ In the
superalgebra case, the charges $q_{k}$ verify in addition to,%
\begin{equation}
q_{k}-q_{k+1}=\pm 1  \label{c2}
\end{equation}%
the super-traceless condition%
\begin{equation}
str\left( \mathbf{\mu }\right) =\sum_{k}q_{k}str\left( \Pi _{i}\right) =0
\label{c1}
\end{equation}%
These constraints allow us to compute these charges in the absence of
branching rules for \textrm{representations of} superalgebras in the
literature.

\textbf{C)} \emph{Super Lax operator }$\mathcal{L}_{\mathbf{R}}^{\mathbf{\mu
}}$\newline
The super L-operator describing the coupling of the representation $\mathbf{R%
}$ and the magnetic coweight $\mathbf{\mu }$\ acting on the superalgebra $%
\mathbf{g}_{\text{\textsc{bf}}}$ as (\ref{dec}), is equal to%
\begin{equation}
\mathcal{L}_{\mathbf{R}}^{\mathbf{\mu }}=e^{X}z^{\mathbf{\mu }}e^{Y}
\label{lax}
\end{equation}%
Later on, it will be simply labeled as $\mathcal{L}^{\mathbf{\mu }}$ since
we will take $\mathbf{R}$ as the fundamental representation for every
symmetry type. The $z^{\mathbf{\mu }}$ follows from the action of $\mathbf{%
\mu }$\ on $\mathbf{R}$ as given by (\ref{cow}). The$\ X$ and $Y$\ are
elements of $\mathbf{N}_{+}$\ and\ $\mathbf{N}_{-}$\ expand \textrm{in
general} like%
\begin{equation}
X=\sum_{i}b^{i}X_{i}+\sum_{\alpha }\beta ^{\alpha }\mathcal{X}_{\alpha
}\qquad ,\qquad Y=\sum_{i}c_{i}Y^{i}+\sum_{\alpha }\gamma _{\alpha }\mathcal{%
Y}^{\alpha }
\end{equation}%
\textrm{where }$X_{i}$\textrm{\ and }$Y^{i}$\textrm{\ are bosonic
generators, and }$\left( \mathcal{X}_{\alpha },\mathcal{Y}^{\alpha }\right) $%
\textrm{\ are fermionic ones. These graded root generators correspond to
roots of} $\mathbf{g}_{\text{\textsc{bf}}}$ that are not contained in $%
\boldsymbol{l}_{\mathbf{\mu }}=\mathfrak{l}\mathbf{_{\mathbf{\mu }}}\oplus
\mathbb{C}\mathbf{\mu }$\textrm{, they} can be realized by using the
following property: For a cutted node corresponding to a graded simple root $%
\widetilde{\beta },$ the root system $\Phi _{\mathbf{g}_{\text{\textsc{bf}}%
}} $ splits as%
\begin{equation}
\Phi _{\mathbf{g}_{\text{\textsc{bf}}}}=\Phi _{\boldsymbol{l}\mathbf{_{%
\mathbf{\mu }}}}\cup \Phi _{\mathbf{N}_{\pm }}\quad ,\quad \Phi _{%
\boldsymbol{l}\mathbf{_{\mathbf{\mu }}}}=\Phi _{\mathfrak{l}\mathbf{_{%
\mathbf{\mu }}}}
\end{equation}%
where $\Phi _{\mathbf{N}_{\pm }}$ contains graded roots in $\Phi _{\mathbf{g}%
_{\text{\textsc{bf}}}}^{\prime }$ that depend on $\widetilde{\beta }$, i.e.%
\begin{eqnarray}
\Phi _{\mathbf{N}_{\pm }} &=&\left \{ \pm \widetilde{\alpha }_{\text{\textsc{%
bf}}}\in \Phi _{\mathbf{g}_{\text{\textsc{bf}}}}^{\prime },\quad \frac{%
\partial \widetilde{\alpha }_{\text{\textsc{bf}}}}{\partial \widetilde{\beta
}}\neq 0\right \} \\
\Phi _{\boldsymbol{l}\mathbf{_{\mathbf{\mu }}}} &=&\left \{ \pm \widetilde{%
\alpha }_{\text{\textsc{bf}}}\in \Phi _{\mathbf{g}_{\text{\textsc{bf}}%
}}^{\prime },\quad \frac{\partial \widetilde{\alpha }_{\text{\textsc{bf}}}}{%
\partial \widetilde{\beta }}=0\right \}
\end{eqnarray}%
The sign of roots in each nilpotent is defined by the condition (\ref{cond}%
), such that $\mathbf{\mu }$\ acts on \textrm{generators} of $\mathbf{N}%
_{\pm }$ with $\pm 1.$\textrm{\ As an illustrative example in the bosonic
linear algebra, the Levi decomposition obtained by the cutting of the node }$%
\widetilde{\alpha }_{3}$\textrm{\ from the Dynkin diagram of} $A_{3}$\textrm{%
\ having the simple roots }$\left \{ \widetilde{\alpha }_{1},\widetilde{%
\alpha }_{2},\widetilde{\alpha }_{3}\right \} $\textrm{\ reads as }$%
sl(4)\rightarrow sl(3)\oplus gl(1).$ \textrm{The root system of }$sl(4)$%
\textrm{\ containing the 12 roots}%
\begin{equation*}
\pm \widetilde{\alpha }_{1},\pm \widetilde{\alpha }_{2},\pm \widetilde{%
\alpha }_{3},\pm \left( \widetilde{\alpha }_{1}+\widetilde{\alpha }%
_{2}\right) ,\pm \left( \widetilde{\alpha }_{2}+\widetilde{\alpha }%
_{3}\right) ,\pm \left( \widetilde{\alpha }_{1}+\widetilde{\alpha }_{2}+%
\widetilde{\alpha }_{3}\right)
\end{equation*}%
\textrm{splits as\ }%
\begin{eqnarray*}
\Phi _{sl_{4}} &=&\Phi _{sl_{3}}+\Phi _{N_{\pm }} \\
\Phi _{sl_{3}} &=&\pm \widetilde{\alpha }_{1},\pm \widetilde{\alpha }%
_{2},\pm \left( \widetilde{\alpha }_{1}+\widetilde{\alpha }_{2}\right) \\
\Phi _{n_{\pm }} &=&\pm \widetilde{\alpha }_{3},\pm \left( \widetilde{\alpha
}_{2}+\widetilde{\alpha }_{3}\right) ,\pm \left( \widetilde{\alpha }_{1}+%
\widetilde{\alpha }_{2}+\widetilde{\alpha }_{3}\right)
\end{eqnarray*}%
\textrm{Here, the 6 roots independent of }$\widetilde{\alpha }_{3}$\textrm{\
correspond to }$\mathfrak{l}_{\mu _{3}}=sl(3),$\textrm{\ and the six roots
depending on }$\widetilde{\alpha }_{3}$\textrm{\ generate the subspaces }$%
n_{\pm }.$\newline
\textrm{Finally, notice that the bosonic coefficients }$\left(
b^{i},c_{i}\right) ,$\textrm{\ and the fermionic }$\left( \beta ^{\alpha
},\gamma _{\alpha }\right) $\textrm{\ form graded oscillators verifying the
Poisson brackets}%
\begin{equation}
\left \{ b^{i},c_{j}\right \} _{PB}=\delta _{j}^{i}\qquad ,\qquad \left \{
\beta ^{\alpha },\gamma _{\lambda }\right \} _{PB}=\delta _{\lambda
}^{\alpha }
\end{equation}%
\textrm{Their quantum versions obey the usual super Heisenberg algebra}.%
\newline
These steps will be used below to complete missing results in literature
concerning integrable superspin chains with underlying symmetries given by
the Lie superalgebras like $B_{\text{\textsc{bf}}}$, $C_{\text{\textsc{bf}}}$
and $D_{\text{\textsc{bf}}}$.\textrm{\ }But before that, we begin by testing
\textrm{this approach by building the oscillator realizations of super
L-operators for the }$sl(m|n)$\ superspin chain, and comparing with
equivalent super matrices in the literature. \textrm{The possible 3-gradings
of the type (\ref{dec}) that we will be using are classified for the family
of complexified basic Lie superalgebras as follows \cite{van}}%
\begin{equation}
\begin{tabular}{|c|c|c|}
\hline
$\mathbf{g}_{\text{\textsc{bf}}}$ & $\mathbf{g}_{0}$ & $\dim \mathbf{g}%
_{-1}=\dim \mathbf{g}_{+1}$ \\ \hline
$A(m|n)$ & $sl(k|l)\oplus sl(m-k|n-l)\oplus gl(1)$ & $(k+l)(m-k+n-l)$ \\
\hline
$B(m|n)$ & $B(m-1|n)\oplus gl(1)$ & $2m+2n-1$ \\ \hline
$C(n)$ & $C_{n-1}\oplus gl(1)$ & $2\left( n-1\right) $ \\ \hline
& $sl(1|n-1)\oplus gl(1)$ & $\frac{n(n+1)}{2}-1$ \\ \hline
$D(m|n)$ & $D(m-1|n)\oplus gl(1)$ & $2\left( m+n-1\right) $ \\ \hline
& $sl(m|n)\oplus gl(1)$ & $\frac{\left( m+n\right) (m+n+1)}{2}-m$ \\ \hline
\end{tabular}
\label{219}
\end{equation}

\section{Super L-operators for all $sl(m|n)$ superspin chains}

\label{sec:3} 
\begin{figure}[h]
\begin{center}
\includegraphics[width=12cm]{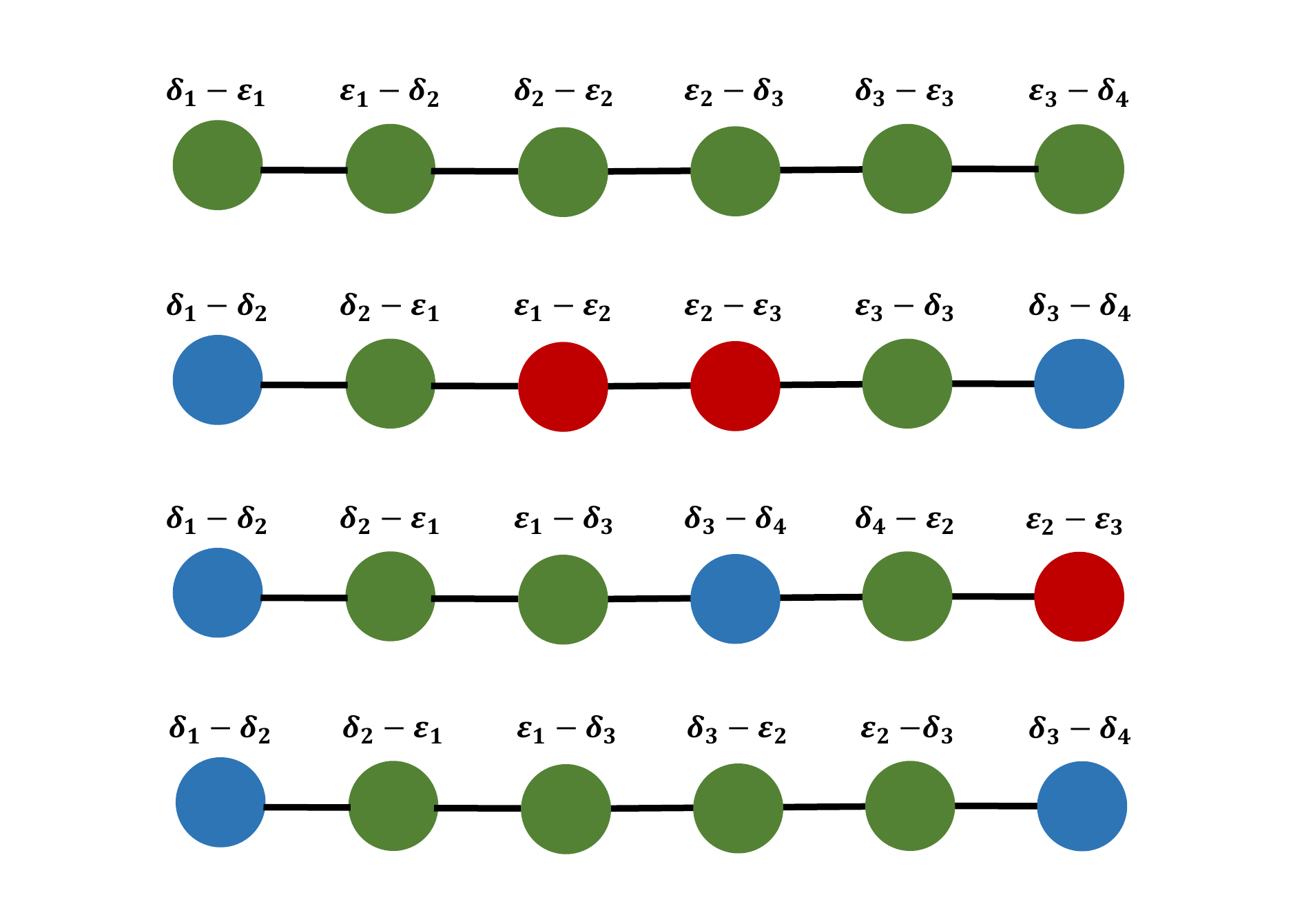}
\end{center}
\par
\vspace{-0.5cm}
\caption{Graded Dynkin diagrams of the $sl(3|4)$ superalgebra. The green
nodes represent fermionic simple roots, blue nodes represent bosonic roots
with $\protect \alpha ^{2}=-2$ and red nodes represent bosonic roots with $%
\protect \alpha ^{2}=2.$}
\label{2G}
\end{figure}

\begin{figure}[h]
\begin{center}
\includegraphics[width=14cm]{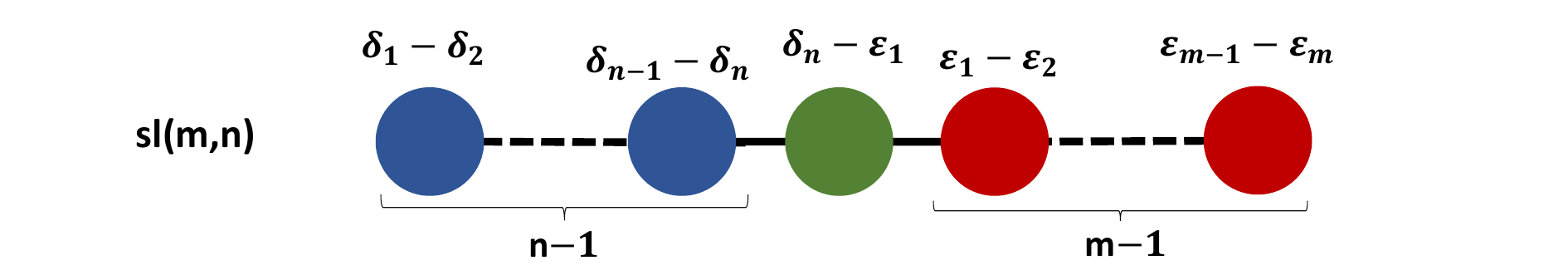}
\end{center}
\par
\vspace{-0.5cm}
\caption{Distinguished Dynkin diagrams for the $sl(m|n)$ superalgebra. The
green node represents the only fermionic node, blue and red nodes are
bosonic.}
\label{am}
\end{figure}

In this section, we apply the formula (\ref{lax}) introduced
in the previous section in order to build super L-operators for the $sl(m|n)$%
\ superspin chain. Thanks to the richness of this A-type supersymmetry, we\
will be able to generate all families of solutions $\mathcal{L}^{\mathbf{\mu
}}$\ labeled by magnetic charges $\mathbf{\mu }$\ of $SL(m|n).$\ These
coweights are in one to one with different nodes of different Dynkin
diagrams $\mathfrak{D}[sl(m|n)]^{(\mathrm{\kappa })}$\ of $sl(m|n)$\ labeled
by positive integers $\mathrm{\kappa }.$ In these regards, recall that
contrary to bosonic Lie algebras $g_{\text{\textsc{bose}}}$\textrm{,} a
superalgebra $\mathbf{g}_{\text{\textsc{bf}}}$\ has several Dynkin diagrams $%
\mathfrak{D}[\mathbf{g}_{\text{\textsc{bf}}}]^{(\mathrm{\kappa })};$ this is
due to the existence of two kinds of fundamental\ unit weight vectors :
bosonic unit \textrm{weights }$\varepsilon _{a}$\textrm{\ }with metric $%
\left \langle \varepsilon _{a},\varepsilon _{b}\right \rangle =\delta _{ab}$%
, and fermionic $\mathrm{\delta }_{\text{\textsc{a}}}$'s with $\left \langle
\mathrm{\delta }_{\text{\textsc{a}}},\mathrm{\delta }_{\text{\textsc{b}}%
}\right \rangle =-\delta _{\text{\textsc{ab}}}$. Hence, the graded simple
roots $\widetilde{\alpha }_{i}$\textrm{\ }have special properties depending
on their realisations, which for $sl(m|n)$\textrm{\ }may be $(i)$ fermionic
of the form
\begin{equation}
\widetilde{\alpha }_{\text{\textsc{a}}a}=\mathrm{\delta }_{\text{\textsc{a}}%
}-\varepsilon _{a}\qquad ,\qquad \widetilde{\alpha }_{\text{\textsc{a}}%
a}^{2}=0
\end{equation}%
or $(ii)$\ bosonic having two possible forms like%
\begin{equation}
\begin{tabular}{lllllll}
$\widetilde{\alpha }_{\text{\textsc{a}}}$ & $=$ & $\mathrm{\delta }_{\text{%
\textsc{a}}}-\mathrm{\delta }_{\text{\textsc{a}}+1}$ & $\qquad ,\qquad $ & $%
\widetilde{\alpha }_{\text{\textsc{a}}}^{2}$ & $=$ & $-2$ \\
$\widetilde{\alpha }_{a}$ & $=$ & $\varepsilon _{a}-\varepsilon _{a+1}$ & $%
\qquad ,\qquad $ & $\widetilde{\alpha }_{a}^{2}$ & $=$ & $+2$%
\end{tabular}%
\end{equation}%
Recall also that, as for bosonic $g_{\text{\textsc{bose}}}$, the set of the
simple roots generate the graded root system $\Phi _{\mathbf{g}_{\text{%
\textsc{bf}}}}\equiv \{ \pm \widetilde{\alpha }_{\text{\textsc{bf}}}\}$; and
because of the three possibilities, we distinguish different types of root
systems for $\mathbf{g}_{\text{\textsc{bf}}}$ labeled by $\mathrm{\kappa }$
and denoted like $\Phi _{\mathbf{g}_{\text{\textsc{bf}}}}^{\left( \mathrm{%
\kappa }\right) }.$ Generally, a superalgebra $\mathbf{g}_{\text{\textsc{bf}}%
}$ has several Dynkin diagrams $\mathfrak{D}[\mathbf{g}_{\text{\textsc{bf}}%
}]^{(\mathrm{\kappa })}$%
\begin{equation}
\mathfrak{D}[\mathbf{g}_{\text{\textsc{bf}}}]^{\left( 1\right) },\quad
\mathfrak{D}[\mathbf{g}_{\text{\textsc{bf}}}]^{\left( 2\right) },\quad
\mathfrak{D}[\mathbf{g}_{\text{\textsc{bf}}}]^{\left( 3\right) },\quad
....\quad \mathfrak{D}[\mathbf{g}_{\text{\textsc{bf}}}]^{(n_{\text{\textsc{bf%
}}})}
\end{equation}%
As an illustration, we give in \textbf{Figure \ref{2G}} examples of\ super
Dynkin diagrams $\mathfrak{D}[sl_{(3|4)}]^{(\mathrm{\kappa })}$ concerning
the $sl(3|4)$ superalgebra. In this Figure, the fermionic simple roots are
represented by green nodes, the bosonic simple roots with $\widetilde{\alpha
}^{2}=-2$ are represented in blue, and the bosonic simple roots with $%
\widetilde{\alpha }^{2}=2$ in red. Recall that for this Lie superalgebra, we
actually have
\begin{equation}
\frac{7!}{4!\times 3!}=35
\end{equation}%
possible super Dynkin diagrams $\mathfrak{D}[sl_{(3|4)}]^{(1)},...,\mathfrak{%
D}[sl_{(3|4)}]^{(35)}$ depending on the ordering of the fundamental unit
weights $\mathrm{\delta }_{\text{\textsc{a}}},\varepsilon _{a}$.
The distinguished super Dynkin diagram having only one fermionic node is
drawn for a general Lie superalgebra $sl(m|n)$ in \textbf{Figure \ref{am}};
the weight basis associated to such diagram is also \textrm{known} as the
distinguished \textrm{basis}.

This multiplicity of Dynkin diagrams for a superalgebra results in different
possible varieties of superspin chains for each symmetry $g_{\text{\textsc{bf%
}}}.$\ For the $sl(m|n)$\ symmetry, we will consider all the possible
superspin chain systems and build the general super Lax operator $\mathcal{L}%
_{sl(m|n)}^{\mathbf{\mu }}$\ associated to a generic node of one of the $%
\left( m+n\right) !/(m!n!)$\ super Dynkin diagrams $\mathfrak{D}%
[sl_{(m|n)}]^{(\mathrm{\kappa })}$\textrm{.} Notice that these general
solutions include those derived in \cite{slmn} using the same approach but
only for the distinguished Dynkin diagram having one fermionic node given by
$\widetilde{\alpha }_{m}=\mathrm{\delta }_{n}-\varepsilon _{1}$ (\textbf{%
Figure \ref{am})}\textrm{.}\newline
In order to proceed for the calculation, we begin by recalling that the Lie
superalgebra $A(m-1|n-1)=sl(m|n)$ \textrm{with} $n\neq m$ has rank $m+n-1$
and $\left( m+n\right) ^{2}-1$ dimensions. It has two sectors: an even sector%
\textrm{\ }%
\begin{equation}
sl(m|n)_{\bar{0}}=sl(m)\oplus sl(n)\oplus gl(1)
\end{equation}%
describing bosons; and an odd sector\textrm{\ }$sl(m|n)_{\bar{1}}$ given by
the module\textrm{\ }$m|\bar{n}\oplus \bar{m}|n$ describing fermions, it
will be denoted below as $2mn$ \textrm{for short.}\newline
The general 3-grading for the $sl(m|n)$\ superalgebra is written as%
\begin{equation}
\begin{tabular}{ccc}
$sl(m|n)$ & $\rightarrow $ & $l_{\mu }\oplus gl(1)\oplus \mathbf{N}%
_{+}\oplus \mathbf{N}_{-}$ \\
$\boldsymbol{l}_{\mathbf{\mu }}$ & $=$ & $sl(k|l)\oplus sl(m-k|n-l)$%
\end{tabular}%
\end{equation}%
with\textrm{\ }$0\leq k\leq m,$\textrm{\ }$0\leq l\leq n,$\ and%
\begin{equation}
\dim \mathbf{N}_{+}=\dim \mathbf{N}_{-}=\left( k+l\right) \left(
m-k+n-l\right)  \label{6}
\end{equation}%
This grading can correspond on the graphical level to the cutting of any of
the $m+n-1$ nodes (bosonic or fermionic) of an arbitrary super Dynkin
diagram of the $(m+n)!/(m!n!)$ possible diagrams $\mathfrak{D}%
[sl_{(m|n)}]^{(\kappa )}.$\ This is a special property of the linear
symmetry where all nodes act like minuscule coweights. Following this
decomposition, the representations of $sl(m|n)$ also get partitioned\textrm{%
. I}n what concerns us, the fundamental representation $\mathbf{m|n}$ of $%
sl(m|n)$ decomposes into irreps of $sl(k|l)$ and $sl(m-k|n-l)$ as follows%
\begin{equation}
\mathbf{m{\LARGE |}n\rightarrow k{\LARGE |}l}_{1+\frac{l-k}{m-n}}\oplus
\left( \mathbf{m-k}\right) \mathbf{\LARGE |}\left( \mathbf{n-l}\right) _{%
\frac{l-k}{m-n}}  \label{5}
\end{equation}%
\textrm{The subscripts refer to the charges of these subspaces with respect
to the }$GL(1)$\textrm{\ corresponding to the cut node, they are calculated
using the conditions (\ref{c2}), (\ref{c1}). }Eq(\ref{5}) can be further
decomposed into four blocks containing $\left( 1\right) $ the fundamental
\underline{$\mathbf{k}$} of $sl(k)$, $\left( 2\right) $ the fundamental
\underline{$\boldsymbol{l}$} of $sl(l)$, $\left( 3\right) $ the \underline{$%
\mathbf{m-k}$} of $sl(m-k)$; and $\left( 4\right) $ the \underline{$\mathbf{%
n-l}$} of $sl(n-l)$. These are respectively represented by the basis states $%
\left \vert a\right \rangle ,$ $\left \vert i\right \rangle ,$ $\left \vert
\alpha \right \rangle ,$ $\left \vert \lambda \right \rangle $ where
\begin{equation}
1\leq a\leq k,\qquad k+1\leq i\leq m,\qquad m+1\leq \alpha \leq m+l,\qquad
m+l+1\leq \lambda \leq m+n
\end{equation}%
We can now write the action of the coweight in terms of the four
corresponding projectors as%
\begin{equation}
\mathbf{\mu }=\left( 1+\frac{l-k}{m-n}\right) \left( \Pi _{\mathbf{k}}+\Pi _{%
\mathbf{l}}\right) +\left( \frac{l-k}{m-n}\right) \left( \Pi _{\mathbf{m-k}%
}+\Pi _{\mathbf{n-l}}\right)
\end{equation}%
with vanishing super trace%
\begin{equation}
str\left( \mathbf{\mu }\right) =\left( \frac{m-n+l-k}{m-n}\right) \left(
k-l\right) +\frac{l-k}{m-n}\left( m-n+l-k\right) =0
\end{equation}%
The nilpotent operators $X$ and $Y$ belonging to $N_{+}$ and $N_{-}$ (\ref{6}%
) are realized by%
\begin{eqnarray}
X &=&b^{ai}\left \vert a\right \rangle \left \langle i\right \vert +\beta
^{a\lambda }\left \vert a\right \rangle \left \langle \lambda \right \vert
+\beta ^{\alpha i}\left \vert \alpha \right \rangle \left \langle i\right
\vert +b^{\alpha \lambda }\left \vert \alpha \right \rangle \left \langle
\lambda \right \vert \\
Y &=&c_{ia}\left \vert i\right \rangle \left \langle a\right \vert +\gamma
_{\lambda a}\left \vert \lambda \right \rangle \left \langle a\right \vert
+\gamma _{i\alpha }\left \vert i\right \rangle \left \langle \alpha \right
\vert +c_{\lambda \alpha }\left \vert \lambda \right \rangle \left \langle
\alpha \right \vert
\end{eqnarray}%
where summation on repeated indices is omitted. The $\left(
b^{ai},c_{ia}\right) $ and $\left( b^{\alpha \lambda },c_{\lambda \alpha
}\right) $ are couples of bosonic harmonic oscillators while $\left( \beta
^{a\lambda },\gamma _{\lambda a}\right) $\ and\ $\left( \beta ^{\alpha
i},\gamma _{i\alpha }\right) $\ form fermioinc\ oscillators. The L-operator
is computed using the nilpotency properties $X^{2}=0$, $Y^{2}=0$ as well as
\begin{equation}
\begin{tabular}{lll}
$X\Pi _{\mathbf{k}}=X\Pi _{\mathbf{l}}=0$ & $,$ & $\Pi _{\mathbf{k}}Y=\Pi _{%
\mathbf{l}}Y=0$ \\
$X\Pi _{\mathbf{m-k}}=X\Pi _{\mathbf{n-l}}=X$ & $,$ & $\Pi _{\mathbf{m-k}%
}Y=\Pi _{\mathbf{n-l}}Y=Y$%
\end{tabular}%
\end{equation}%
It expands as%
\begin{eqnarray}
\mathcal{L}_{sl_{m|n}}^{\mathbf{\mu }} &=&z^{1+\frac{l-k}{m-n}}\Pi _{\mathbf{%
k}}+z^{1+\frac{l-k}{m-n}}\Pi _{\boldsymbol{l}}+z^{\frac{l-k}{m-n}}\Pi _{%
\mathbf{m-k}}+z^{\frac{l-k}{m-n}}\Pi _{\mathbf{n-}\boldsymbol{l}} \\
&&+X\left( z^{\frac{l-k}{m-n}}\Pi _{\mathbf{m-k}}+z^{\frac{l-k}{m-n}}\Pi _{%
\mathbf{n-}\boldsymbol{l}}\right) +\left( z^{\frac{l-k}{m-n}}\Pi _{\mathbf{%
m-k}}+z^{\frac{l-k}{m-n}}\Pi _{\mathbf{n-}\boldsymbol{l}}\right) Y  \notag \\
&&+X\left( z^{\frac{l-k}{m-n}}\Pi _{\mathbf{m-k}}+z^{\frac{l-k}{m-n}}\Pi _{%
\mathbf{n-}\boldsymbol{l}}\right) Y  \notag
\end{eqnarray}%
yielding the matrix form%
\begin{equation}
\mathcal{L}_{sl_{m|n}}^{\mathbf{\mu }}=z^{h}\left(
\begin{array}{cccc}
z\delta _{b}^{a}+\left( b^{ai}c_{ib}+\beta ^{a\lambda }\gamma _{\lambda
b}\right) & \left( b^{ai}\gamma _{i\alpha }+\beta ^{a\lambda }c_{\lambda
\alpha }\right) & b^{aj}\delta _{ji} & \beta ^{a\rho }\delta _{\rho \lambda }
\\
\left( \beta ^{\alpha i}c_{ib}+b^{\alpha \lambda }\gamma _{\lambda b}\right)
& z\delta _{\eta }^{\alpha }+\left( \beta ^{\alpha i}\gamma _{i\eta
}+b^{\alpha \lambda }c_{\lambda \eta }\right) & \beta ^{\alpha j}\delta _{ji}
& b^{\alpha \rho }\delta _{\rho \lambda } \\
c_{ia} & \gamma _{i\alpha } & \delta _{i}^{j} & 0 \\
\gamma _{\lambda a} & c_{\lambda \alpha } & 0 & \delta _{\lambda }^{\rho }%
\end{array}%
\right)  \label{20}
\end{equation}%
where we have set $h=\frac{l-k}{m-n}.$ This matrix is in agreement with the
general solution obtained in the superspin chain literature; see eq(2.20) in
\textrm{\cite{FRC}}. The special families of solutions corresponding to the
nodes of the distinguished Dynkin diagram are calculated in details in \cite%
{slmn}, where the particular Lax matrix with purely fermionic oscillators is
obtained \textrm{by considering} the only fermionic node of the
distinguished diagram.

\section{Super L-operators of $B(m|n)$ type}

\label{sec:4} 
\begin{figure}[tbph]
\begin{center}
\includegraphics[width=14cm]{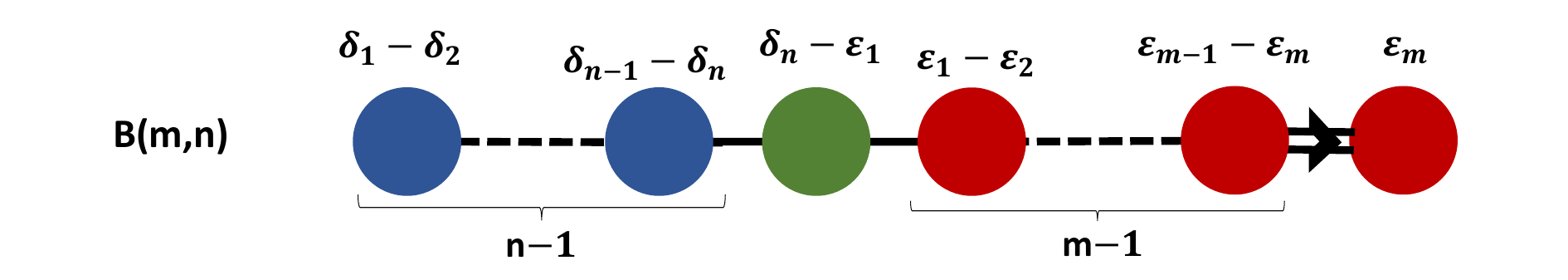}
\end{center}
\par
\vspace{-0.5cm}
\caption{{}Distinguished Dynkin diagram of the $B(m|n)$ superalgebra having
one fermionic simple root in Green color. }
\label{B1}
\end{figure}

In this section, we study the B$_{\text{\textsc{bf}}}$-family
of orthosymplectic integrable superspin chains with internal symmetry given
by the Lie superalgebra series%
\begin{equation}
B(m|n)=osp(2m+1|2n),\qquad m,n>0
\end{equation}%
We focus on the family of distinguished superspin $B(m|n)$ chain associated
to the Distinguished Dynkin diagram, and calculate\textrm{\ the Lax operator
}$\mathcal{L}_{B_{m|n}}^{\mathbf{\mu }_{n+1}}$\textrm{\ }by using the
3-grading of the orthosymplectic $B(m|n)$\ in the formula (\ref{lax}).%
\textrm{\newline
}To begin, recall that the $B(m|n)$ superalgebra is a $\mathbb{Z}_{2}$-
graded Lie algebra of rank $r\left( B_{m|n}\right) =m+n,$ and dimension $%
\dim B_{m|n}=2(m+n)^{2}+m+3n.$ \textrm{I}t splits like $B(m|n)_{\bar{0}%
}\oplus B(m|n)_{\bar{1}}$ with even part as,%
\begin{equation}
\begin{tabular}{lll}
$B(m|n)_{\bar{0}}$ & $=$ & $B_{m}\oplus C_{n}$ \\
& $\simeq $ & $so(2m+1)\oplus sp\left( 2n\right) $%
\end{tabular}%
\end{equation}%
and odd part $B(m|n)_{\bar{1}}$ generated by the bi-fundamental
representation $(2m+1,2n)$ of $so(2m+1)\oplus sp\left( 2n\right) .$ The root
system $\Phi _{B_{m|n}}$ of the Lie superalgebra $B(m|n)$ has $%
2(m+n)^{2}+2n\ $elements; it does also split into an even part $\Phi _{\bar{0%
}}$ and an odd part $\Phi _{\bar{1}}$. By using the unit bosonic weight
vectors $\left \{ \varepsilon _{a}\right \} _{1\leq a\leq m}$ and the
fermionic $\left \{ \mathrm{\delta }_{\text{\textsc{a}}}\right \} _{1\leq
\text{\textsc{a}}\leq n},$ the content of $\Phi _{\bar{0}}$ reads as%
\begin{equation}
\begin{tabular}{lllllll}
$\Phi _{\bar{0}}$ & $:$ & $\pm \left( \varepsilon _{a}\pm \varepsilon
_{b}\right) $ & $,$ & $\pm \varepsilon _{a}$ & $,\qquad $ & $a\neq b=1,...,m$
\\
&  & $\pm \left( \mathrm{\delta }_{\text{\textsc{a}}}\pm \mathrm{\delta }_{%
\text{\textsc{b}}}\right) $ & $,$ & $\pm 2\mathrm{\delta }_{\text{\textsc{a}}%
}$ & $,$ & \textsc{a}$\neq $\textsc{b}$=1,...,n$%
\end{tabular}%
\end{equation}%
with cardinal $\left \vert \Phi _{\bar{0}}\right \vert =2m^{2}+2n^{2}$, and
the roots of $\Phi _{\bar{1}}$ read like%
\begin{equation}
\Phi _{\bar{1}}:\pm \mathrm{\delta }_{\text{\textsc{a}}},\qquad \pm \left(
\varepsilon _{a}\pm \mathrm{\delta }_{\text{\textsc{a}}}\right)
\end{equation}%
with $\left \vert \Phi _{\bar{1}}\right \vert =2n+2mn$. Given the set $\Phi
_{B_{m|n}}$, a remarkable simple root basis generating it is given by the
distinguished basis $(\widetilde{\beta }_{\text{\textsc{a}}},\widetilde{%
\gamma }\mathrm{,}\widetilde{\alpha }_{a})$ having one fermionic root $%
\widetilde{\gamma }=\mathrm{\delta }_{n}-\varepsilon _{1}$ with length $%
\widetilde{\gamma }^{2}=0$ ; and $m+n-1$ bosonic ones as%
\begin{equation}
\widetilde{\beta }_{\text{\textsc{a}}}=\mathrm{\delta }_{\text{\textsc{a}}}-%
\mathrm{\delta }_{\text{\textsc{a+1}}}\qquad ,\qquad \widetilde{\alpha }%
_{a}=\varepsilon _{a}-\varepsilon _{a+1}\qquad ,\qquad \widetilde{\alpha }%
_{m}=\varepsilon _{m}
\end{equation}%
with $\widetilde{\beta }_{\text{\textsc{a}}}^{2}=-2$ and $\widetilde{\alpha }%
_{a}^{2}=2$. The distinguished basis is characterised by the following
ordering of the fundamental unit weight vectors%
\begin{equation}
\mathrm{\delta }_{1},\quad \mathrm{\delta }_{2},\quad ...\quad \mathrm{%
\delta }_{n-1},\quad \mathrm{\delta }_{n};\quad \varepsilon _{1},\quad
\varepsilon _{2},\quad ...\quad \varepsilon _{m-1},\quad \varepsilon _{m}
\label{order}
\end{equation}%
for which the super Cartan matrix has the entries
\begin{equation}
C_{B_{m|n}}^{dist}=\left(
\begin{array}{ccccccccc}
-2 & 1 & \cdots &  &  &  &  &  &  \\
1 & -2 & 1 &  &  &  &  &  &  \\
& \ddots & \ddots &  &  &  &  &  &  \\
&  & 1 & -2 & 1 &  &  &  &  \\
&  &  & 1 & 0 & -1 &  &  &  \\
&  & \cdots &  & -1 & 2 & -1 &  &  \\
&  &  &  &  & \ddots & \ddots & \ddots &  \\
&  &  &  &  &  & -1 & 2 & -1 \\
&  &  &  &  &  &  & -1 & 1%
\end{array}%
\right)
\end{equation}%
with only one zero on the diagonal. The distinguished Dynkin diagram
corresponding to this matrix is given by \textbf{Figure \ref{B1}}.
Notice that due to the $Z_{2}$- grading, we can actually distinguish\textrm{%
\ }%
\begin{equation}
N_{n,m}^{\mathrm{B}_{\text{\textsc{bf}}}}=\frac{\left( n+m-1\right) !}{%
n!\times \left( m-1\right) !}+\frac{\left( n+m-1\right) !}{m!\times \left(
n-1\right) !}
\end{equation}%
types of super Dynkin diagrams $\mathfrak{D}{\small [B}_{{\small m|n}}%
{\small ]}$\textrm{\ }having $m+n$\textrm{\ }nodes represented by different
forms of graded simple roots\textrm{\ }$\tilde{\alpha}.$\newline
In what follows, we will consider the distinguished basis and construct the
Lax operator for the corresponding super $B$- chain. In this regard, recall
that by distinguished orthosymplectic super chain of $B$- type, we mean the
two following: \newline
$\left( \mathbf{1}\right) $ an integrable superspin $B_{m|n}$ chain made of
\textquotedblleft super atoms\textquotedblright \ arranged along a straight
line; and realised in the CS theory (\ref{7}) in terms of a set of parallel
super Wilson lines traversed by a horizontal 't Hooft line. Such a
realisation by topological defects looks like the one investigated in \cite%
{slmn} for the case of $sl(m|n)$ superspin chain; the main difference is
that here the super \textrm{spins} are of $B$-type instead of $A$-type.
\newline
$\left( \mathbf{2}\right) $ The vertical super Wilson lines are run by
graded quantum states $(\mathrm{\delta }_{\text{\textsc{a}}}\mathbf{|}%
\varepsilon _{a})$ ordered as in eq(\ref{order}) and interpreted in terms of
the distinguished Dynkin diagram $\mathfrak{DD}\left[ B_{m|n}\right] $ given
by \textbf{Figure \ref{B1}}. \newline
\textrm{Given }this orthosymplectic superspin chain configuration and the
associated super Dynkin diagram characterising each super Wilson line (a
super-atom), we can calculate the distinguished super L-operator $\mathcal{L}%
_{B_{n|m}}^{\mathbf{\mu }_{n+1}}$ following from the 3-grading \cite{van}

\begin{equation}
B(m|n)\rightarrow A_{1}\oplus B(m-1|n)\oplus \mathbf{N}_{+}\oplus \mathbf{N}%
_{-}
\end{equation}%
or equivalently%
\begin{equation}
osp(2m+1|2n)\rightarrow gl(1)\oplus osp(2m-1|2n)\oplus \mathbf{N}_{+}\oplus
\mathbf{N}_{-}  \label{osp}
\end{equation}%
with nilpotents
\begin{equation}
\mathbf{N}_{\pm }=2m+2n-1  \label{osp1}
\end{equation}%
The fundamental representation decomposes in this case as
\begin{equation}
\mathbf{2m+1|2n}\rightarrow \mathbf{2}\oplus (\mathbf{2m-1|2n)}
\end{equation}%
We further use the reducibility $\mathbf{2}_{0}=\mathbf{1}_{+}\oplus \mathbf{%
1}_{-}$\ to reveal the Levi-like charges under the $GL(1)$\ of the cut node.
Thus, we can rewrite the above decomposition as follows%
\begin{equation}
\left( \mathbf{2m+1|2n}\right) _{0}\rightarrow \mathbf{1}_{+}\oplus \left(
\mathbf{2m-1|2n}\right) _{0}\oplus \mathbf{1}_{-}  \label{8}
\end{equation}%
Now, in order to realize the components of the super L-operator, we work in
the graded basis%
\begin{equation}
\left \{
\begin{array}{c}
\left \vert \mathbf{+}\right \rangle \\
\left \vert i\right \rangle \\
\left \vert \mathbf{-}\right \rangle \\
\left \vert \alpha \right \rangle%
\end{array}%
\right \}  \label{osp2}
\end{equation}%
where $\left \vert \mathbf{\pm }\right \rangle $ refer to the two singlets $%
\mathbf{1}_{\mathbf{\pm }}$, the states $\left \vert i\right \rangle $\ with
$1\leq i\leq 2m-1$\ correspond\ to the $\mathbf{2m-1}$\ and the fermionic $%
\left \vert \alpha \right \rangle $\ with $1\leq \alpha \leq 2n$\ to the
symplectic vector $\mathbf{2n.}$ In this basis, the coweight $\mathbf{\mu }%
_{n+1}$\ associated to the decomposition\ (\ref{osp}) is written as%
\begin{equation}
\mathbf{\mu }_{n+1}=\varrho _{+}+q_{1}\Pi _{1}-\varrho _{-}+q_{2}\Pi _{2}
\label{mu}
\end{equation}%
where $q_{1}=q_{2}=0$ and the projectors on the subspaces of the fundamental
representation are defined by%
\begin{equation}
\varrho _{\pm }=\left \vert \pm \right \rangle \left \langle \pm \right
\vert ,\qquad \Pi _{1}=\sum_{i=1}^{2m-1}\left \vert i\right \rangle \left
\langle i\right \vert ,\qquad \Pi _{2}=\sum_{\alpha =1}^{2n}\left \vert
\alpha \right \rangle \left \langle \alpha \right \vert
\end{equation}%
The $2(2m+2n-1)$ elements of the $N_{\pm }$ (eq.\ref{osp1}) are realized
here as%
\begin{equation}
X=b^{i}X_{i}+\beta ^{\alpha }\mathcal{X}_{\alpha }\in N_{+}\qquad ,\qquad
Y=c_{i}Y^{i}+\gamma _{\alpha }\mathcal{Y}^{\alpha }\in N_{-}
\end{equation}%
where $\left( b^{i},c_{i}\right) $ are bosonic oscillators and $\left( \beta
^{\alpha },\gamma _{\alpha }\right) $ are fermionic ones; the generators%
\begin{eqnarray}
X_{i} &=&\left \vert \mathbf{+}\right \rangle \left \langle i\right \vert
-\left \vert i\right \rangle \left \langle -\right \vert \qquad ;\qquad
\mathcal{X}_{\alpha }=\left \vert \mathbf{+}\right \rangle \left \langle
\alpha \right \vert -\left \vert \alpha \right \rangle \left \langle -\right
\vert  \label{9} \\
Y^{i} &=&\left \vert i\right \rangle \left \langle \mathbf{+}\right \vert
-\left \vert -\right \rangle \left \langle i\right \vert \qquad ;\qquad
\mathcal{Y}^{\alpha }=\left \vert \alpha \right \rangle \left \langle
\mathbf{+}\right \vert -\left \vert -\right \rangle \left \langle \alpha
\right \vert  \notag
\end{eqnarray}%
verify the Levi-like constraints (\ref{cond}) $\left[ \mathbf{\mu ,}X\right]
=X,$ $\left[ \mathbf{\mu ,}Y\right] =-Y,$ and $\left[ X_{i}\mathbf{,}Y^{i}%
\right] =\left[ \mathcal{X}_{\alpha }\mathbf{,}\mathcal{Y}^{\alpha }\right] =%
\mathbb{C}\mathbf{\mu .}$ To substitute these realizations into the
L-operator formula, we calculate%
\begin{equation}
\begin{tabular}{lll}
$X^{2}$ & $=$ & $-\mathbf{b}^{2}\left \vert \mathbf{+}\right \rangle \left
\langle -\right \vert -\boldsymbol{\beta }^{2}\left \vert \mathbf{+}\right
\rangle \left \langle -\right \vert $ \\
$Y^{2}$ & $=$ & $-\mathbf{c}^{2}\left \vert -\right \rangle \left \langle
\mathbf{+}\right \vert -\boldsymbol{\gamma }^{2}\left \vert -\right \rangle
\left \langle \mathbf{+}\right \vert $%
\end{tabular}%
\end{equation}%
where we set
\begin{equation}
\begin{tabular}{lll}
$\mathbf{b}^{2}$ & $=$ & $b^{i}\delta _{ij}b^{j}$ \\
$\mathbf{c}^{2}$ & $=$ & $c_{i}\delta ^{ij}c_{j}$%
\end{tabular}%
\end{equation}%
\ and
\begin{equation}
\begin{tabular}{lll}
$\boldsymbol{\beta }^{2}$ & $=$ & $\beta ^{\alpha }\beta _{\alpha }$ \\
$\boldsymbol{\gamma }^{2}$ & $=$ & $\gamma _{\alpha }\gamma ^{\alpha }$%
\end{tabular}%
\end{equation}%
The nilpotency properties $X^{3}=0$ and $Y^{3}=0$ yield $e^{X}=1+X+\frac{1}{2%
}X^{2}$ and $e^{Y}=1+Y+\frac{1}{2}Y^{2}.$ By substituting (\ref{mu}) into $%
z^{\mathbf{\mu }_{n+1}}$, we also have
\begin{equation}
z^{\mathbf{\mu }_{n+1}}=z\varrho _{+}+\Pi _{1}+z^{-1}\varrho _{-}+\Pi _{2}
\end{equation}%
So, the expression of the super L-operator reads as follows%
\begin{equation}
\mathcal{L}_{B_{n|m}}^{\mathbf{\mu }_{n+1}}=\left( 1+X+\frac{1}{2}%
X^{2}\right) z^{\mathbf{\mu }_{n+1}}\left( 1+Y+\frac{1}{2}Y^{2}\right)
\end{equation}%
and expands like%
\begin{equation}
\begin{tabular}{lll}
$\mathcal{L}_{B_{n|m}}^{\mathbf{\mu }_{n+1}}$ & $=$ & $z\varrho _{+}+\Pi
_{1}+z^{-1}\varrho _{-}+\Pi _{2}+X\left( \Pi _{1}+z^{-1}\varrho _{-}+\Pi
_{2}\right) +$ \\
&  & $\left( \Pi _{1}+z^{-1}\varrho _{-}+\Pi _{2}\right) Y+X\left( \Pi
_{1}+z^{-1}\varrho _{-}+\Pi _{2}\right) Y$ \\
&  & $+\frac{1}{2}X^{2}z^{-1}\varrho _{-}+\frac{1}{2}z^{-1}\varrho _{-}Y^{2}+%
\frac{1}{2}X^{2}z^{-1}\varrho _{-}Y$ \\
&  & $+\frac{1}{2}Xz^{-1}\varrho _{-}Y^{2}+\frac{1}{4}X^{2}z^{-1}\varrho
_{-}Y^{2}$%
\end{tabular}%
\end{equation}%
where we have used the properties%
\begin{equation}
\begin{tabular}{lll}
$X\varrho _{+}$ & $=\varrho _{+}Y$ & $=0$ \\
$X^{2}\varrho _{+}$ & $=X^{2}\Pi _{1}$ & $=0$ \\
$\varrho _{+}Y^{2}$ & $=\Pi _{1}Y^{2}$ & $=0$ \\
$X^{2}\Pi _{2}$ & $=\Pi _{2}Y^{2}$ & $=0$%
\end{tabular}%
\end{equation}%
\ In the projector basis introduced before, the super L-operator can be
written in matrix language as%
\begin{equation}
\mathcal{L}_{B_{m|n}}^{\mathbf{\mu }_{n+1}}=\left(
\begin{array}{cccc}
\varrho _{+}L\varrho _{+} & \varrho _{+}L\Pi _{1} & \varrho _{+}L\varrho _{-}
& \varrho _{+}L\Pi _{2} \\
\Pi _{1}L\varrho _{+} & \Pi _{1}L\Pi _{1} & \Pi _{1}L\varrho _{-} & \Pi
_{1}L\Pi _{2} \\
\varrho _{-}L\varrho _{+} & \varrho _{-}L\Pi _{1} & \varrho _{-}L\varrho _{-}
& \varrho _{-}L\Pi _{2} \\
\Pi _{2}L\varrho _{+} & \Pi _{2}L\Pi _{1} & \Pi _{2}L\varrho _{-} & \Pi
_{2}L\Pi _{2}%
\end{array}%
\right)  \label{21}
\end{equation}%
where the various blocks are given in terms of oscillators of the 't Hooft
line phase space by%
\begin{equation}
\begin{array}{lll}
\varrho _{+}L\varrho _{+} & = & z+(b^{i}\delta _{i}^{j}c_{j}+\beta ^{\alpha
}\delta _{\alpha }^{\beta }\gamma _{\beta })+\frac{1}{4}z^{-1}\left( \mathbf{%
b}^{2}+\boldsymbol{\beta }^{2}\right) \left( \mathbf{c}^{2}+\boldsymbol{%
\gamma }^{2}\right) \\
\varrho _{+}L\Pi _{1} & = & b^{i}+\frac{1}{2}z^{-1}\left( \mathbf{b}^{2}+%
\boldsymbol{\beta }^{2}\right) c_{i} \\
\varrho _{+}L\varrho _{-} & = & -\frac{1}{2}z^{-1}\left( \mathbf{b}^{2}+%
\boldsymbol{\beta }^{2}\right) \\
\varrho _{+}L\Pi _{2} & = & b^{\alpha }+\frac{1}{2}z^{-1}\left( \mathbf{b}%
^{2}+\boldsymbol{\beta }^{2}\right) \gamma _{\alpha }%
\end{array}
\label{210}
\end{equation}%
and%
\begin{equation}
\begin{tabular}{lll}
$\Pi _{1}L\varrho _{+}$ & $=$ & $c_{i}+\frac{1}{2}z^{-1}b^{i}\left( \mathbf{c%
}^{2}+\boldsymbol{\gamma }^{2}\right) $ \\
$\Pi _{1}L\Pi _{1}$ & $=$ & $\delta _{j}^{i}+z^{-1}b^{i}c_{j}$ \\
$\Pi _{1}L\varrho _{-}$ & $=$ & $-z^{-1}b^{i}$ \\
$\Pi _{1}L\Pi _{2}$ & $=$ & $z^{-1}b^{i}\gamma _{\beta }$%
\end{tabular}
\label{211}
\end{equation}%
and%
\begin{equation}
\begin{tabular}{lll}
$\varrho _{-}L\varrho _{+}$ & $=$ & $-\frac{1}{2}z^{-1}\left( \mathbf{c}^{2}+%
\boldsymbol{\gamma }^{2}\right) $ \\
$\varrho _{-}L\Pi _{1}$ & $=$ & $-z^{-1}c_{i}$ \\
$\varrho _{-}L\varrho _{-}$ & $=$ & $z^{-1}$ \\
$\varrho _{-}L\Pi _{2}$ & $=$ & $-z^{-1}\gamma _{\alpha }$%
\end{tabular}
\label{212}
\end{equation}%
as well as
\begin{equation}
\begin{array}{lll}
\Pi _{2}L\varrho _{+} & = & \gamma _{\alpha }+\frac{1}{2}z^{-1}\beta
^{\alpha }\left( \mathbf{c}^{2}+\boldsymbol{\gamma }^{2}\right) \\
\Pi _{2}L\Pi _{1} & = & z^{-1}\beta ^{\alpha }c_{j} \\
\Pi _{2}L\varrho _{-} & = & -z^{-1}\beta ^{\alpha } \\
\Pi _{2}L\Pi _{2} & = & \delta _{\alpha }^{\beta }+\beta ^{\alpha }\delta
_{\alpha }^{\beta }\gamma _{\beta }%
\end{array}
\label{213}
\end{equation}

\section{Super L-operators of $C(n)$ type}

\label{sec:5} 
\begin{figure}[tbph]
\begin{center}
\includegraphics[width=14cm]{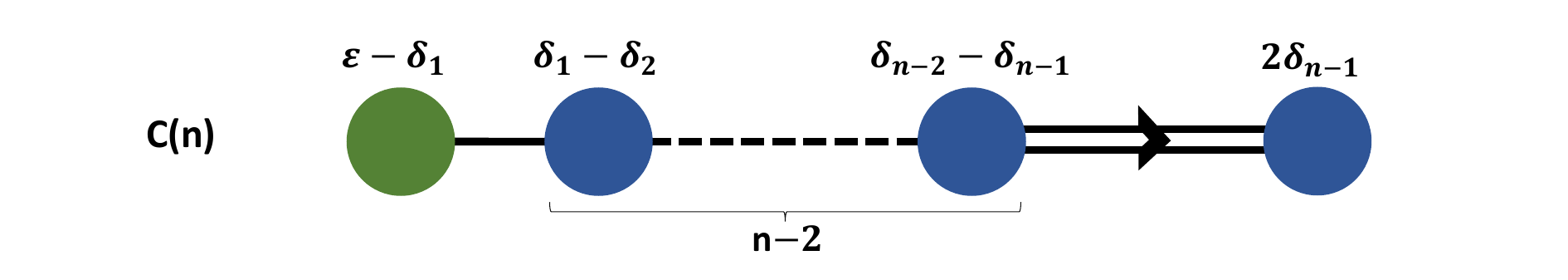}
\end{center}
\par
\vspace{-0.5cm}
\caption{{}Distinguished Dynkin diagram of the $C(n)$ superalgebra, the only
fermionic node is represented in green$.$}
\label{cn}
\end{figure}

Now, by considering the 4D Chern-Simons gauge theory with $%
OSP(2|2n-2)$ symmetry and super Wilson and 't Hooft line defects
implemented, we can build the corresponding $osp(2|2n-2)$\ superspin chain
and compute the RLL solutions in a similar way as before.\newline
The superalgebra $D(1|n-1)=osp(2|2n-2)$ with $n>1$ is labelled as $C(n)$ due
to its even part equal to
\begin{equation}
gl(1)\oplus C_{n-1}=so(2)\oplus sp(2n-2)
\end{equation}%
its odd part is given by $(2n-2)\oplus (2n-2).$ It has rank $n$ and its
dimension is equal to $2n^{2}+n-2.$ The super Cartan matrix reads for the
distinguished basis as follows%
\begin{equation}
C_{C(n)}^{dist}=\left(
\begin{array}{ccccc}
0 & -1 &  &  &  \\
-1 & 2 & -1 &  &  \\
& \ddots & \ddots & \ddots &  \\
&  & -1 & 2 & -2 \\
&  &  & -2 & 4%
\end{array}%
\right)
\end{equation}%
The associated distinguished super Dynkin diagram is depicted in \textbf{%
Figure \ref{cn}}.
For this supersymmetry, we have two possible 3-gradings of the form (\ref%
{dec}); we can therefore construct two super Lax matrices solving the RLL
equation for the distinguished superspin chain of C-type. In fact, the first
one is associated to the only fermionic node $\widetilde{\alpha }_{1}=\delta
-\varepsilon _{1}$\ and the second one to the bosonic node $\widetilde{%
\alpha }_{n}=2\varepsilon _{n-1}.$\ We will begin by working out the first
L-operator $\mathcal{L}_{C(n)}^{\mathbf{\mu }_{1}}$\ that we will label as
fermionic since it only contains fermionic oscillators.

\subsection{The super L-operator $\mathcal{L}_{C(n)}^{\mathbf{\protect \mu }%
_{1}}$}

The first 3-grading for the Lie superalgebra $C(n)$\ is given by%
\begin{equation}
osp(2|2n-2)\quad \rightarrow \quad gl(1)\oplus sp(2n-2)\oplus N_{+}\oplus
N_{-}
\end{equation}%
where the $\boldsymbol{l}_{\mathbf{\mu }_{1}}$ is identified with the
bosonic subalgebra $sp(2n-2)$ and%
\begin{equation}
N_{\pm }=2\left( n-1\right)
\end{equation}%
The dimensions of the $osp(2|2n-2)$\ split like%
\begin{equation}
\begin{tabular}{lll}
$2n^{2}+n-2$ & $\quad \rightarrow \quad $ & $1$ $\  \oplus $ $\ (n-1)\left[
2(n-1)+1\right] \  \  \oplus $ \\
&  & $\  \ N_{+}$ $\  \oplus $ $\ N_{-}$%
\end{tabular}%
\end{equation}%
This decomposition is actually obtained from the distinguished Dynkin
diagram by the cutting of the fermionic node $\mu _{1}$\ dual to $\widetilde{%
\alpha }_{1}$. In what concerns us here, the fundamental representation is
decomposed to representations of the $so(2)\oplus sp(2n-2)$\ as
\begin{equation}
\left( \mathbf{2|2n-2}\right) \quad \rightarrow \quad \mathbf{2}_{0}\oplus
\left( \mathbf{2n-2}\right) _{0}
\end{equation}%
Here as well,\textrm{\ we s}plit the representation $\mathbf{2}_{0}$ into
two singlets $\mathbf{1}_{+}\oplus \mathbf{1}_{-}$\ carrying opposite
charges under the coweight, we work in the basis decomposed as%
\begin{equation}
\left \vert 2n\right \rangle \rightarrow \left \vert +\right \rangle \oplus
\left \vert i\right \rangle \oplus \left \vert -\right \rangle
\end{equation}%
where the states $\left \vert i\right \rangle $ with $1\leq i\leq 2n-2$
correspond to the subspace $\mathbf{2n-2}$. In the same way as before, we
write%
\begin{equation}
\mathbf{\mu }_{1}=\varrho _{+}+q\Pi -\varrho _{-}
\end{equation}%
where $q=0$, $\varrho _{\pm }=\left \vert \pm \right \rangle \left \langle
\pm \right \vert $ and $\Pi =\sum_{i=1}^{2n-2}\left \vert i\right \rangle
\left \langle i\right \vert .$ The $X$ and $Y$ matrices are realized here as%
\begin{equation}
\begin{tabular}{lllll}
$X$ & $=$ & $b^{i}X_{i}$ & $\in $ & $N_{+}$ \\
$Y$ & $=$ & $c_{i}Y^{i}$ & $\in $ & $N_{-}$%
\end{tabular}%
\end{equation}%
with generators%
\begin{equation}
\begin{tabular}{lll}
$X_{i}$ & $=$ & $\left \vert \mathbf{+}\right \rangle \left \langle i\right
\vert -\left \vert i\right \rangle \left \langle -\right \vert $ \\
$Y^{i}$ & $=$ & $\left \vert i\right \rangle \left \langle \mathbf{+}\right
\vert -\left \vert -\right \rangle \left \langle i\right \vert $%
\end{tabular}%
\end{equation}%
such that
\begin{equation}
\left[ \mathbf{\mu }_{1}\mathbf{,}X_{i}\right] =X_{i},\qquad \left[ \mathbf{%
\mu }_{1}\mathbf{,}Y^{i}\right] =-Y^{i},\qquad \left[ X_{i}\mathbf{,}Y^{j}%
\right] =\delta _{i}^{j}\mathbb{C}\mathbf{\mu }_{1}
\end{equation}%
\ We calculate%
\begin{equation}
\begin{tabular}{lllllll}
$X^{2}$ & $=$ & $-\mathbf{b}^{2}\left \vert \mathbf{+}\right \rangle \left
\langle -\right \vert $ & $\qquad ,\qquad $ & $X^{3}$ & $=$ & $0$ \\
$Y^{2}$ & $=$ & $-\mathbf{c}^{2}\left \vert -\right \rangle \left \langle
\mathbf{+}\right \vert $ & $\qquad ,\qquad $ & $Y^{3}$ & $=$ & $0$%
\end{tabular}%
\end{equation}%
where we have set%
\begin{equation}
\mathbf{b}^{2}=b^{i}\delta _{ij}b^{j}\qquad ,\qquad \mathbf{c}%
^{2}=c_{i}\delta ^{ij}c_{j}
\end{equation}%
Eventually, we have%
\begin{equation}
e^{X}=1+X+\frac{1}{2}X^{2}\qquad ,\qquad e^{Y}=1+Y+\frac{1}{2}Y^{2}
\end{equation}%
The L operator formula (\ref{lax}) along with properties $X\varrho
_{+}=\varrho _{+}Y=0$, $X^{2}\varrho _{+}=X^{2}\Pi =\varrho _{+}Y^{2}=\Pi
Y^{2}=0,$\ lead to%
\begin{equation}
\mathcal{L}_{C(n)}^{\mathbf{\mu }_{1}}=\left( 1+X+\frac{1}{2}X^{2}\right)
\left( z\varrho _{+}+\Pi +z^{-1}\varrho _{-}\right) \left( 1+Y+\frac{1}{2}%
Y^{2}\right)
\end{equation}%
expanding like%
\begin{equation}
\begin{tabular}{lll}
$\mathcal{L}_{C(n)}^{\mathbf{\mu }_{1}}$ & $=$ & $z\varrho _{+}+\Pi
+z^{-1}\varrho _{-}+X\left( \Pi +z^{-1}\varrho _{-}\right) +$ \\
&  & $\left( \Pi +z^{-1}\varrho _{-}\right) Y+X\left( \Pi +z^{-1}\varrho
_{-}\right) Y+$ \\
&  & $\frac{1}{2}X^{2}z^{-1}\varrho _{-}+\frac{1}{2}z^{-1}\varrho _{-}Y^{2}+%
\frac{1}{2}X^{2}z^{-1}\varrho _{-}Y+$ \\
&  & $\frac{1}{2}Xz^{-1}\varrho _{-}Y^{2}+\frac{1}{4}X^{2}z^{-1}\varrho
_{-}Y^{2}$%
\end{tabular}%
\end{equation}%
and yielding%
\begin{equation}
\mathcal{L}_{C(n)}^{\mathbf{\mu }_{1}}=\left(
\begin{array}{ccc}
z+b^{i}c_{i}+\frac{1}{4}z^{-1}\mathbf{b}^{2}\mathbf{c}^{2} & b^{i}+\frac{1}{2%
}z^{-1}\mathbf{b}^{2}c_{i} & -\frac{1}{2}z^{-1}\mathbf{b}^{2} \\
c_{i}+\frac{1}{2}z^{-1}\mathbf{c}^{2}b^{i} & \delta _{j}^{i}+z^{-1}b^{i}c_{i}
& -z^{-1}b^{i} \\
-\frac{1}{2}z^{-1}\mathbf{c}^{2} & -z^{-1}c_{i} & z^{-1}%
\end{array}%
\right)  \label{22}
\end{equation}%
This L-operator carries only bosonic oscillator degrees of freedom, which is
expected since the cut node corresponds to the \textrm{only} fermionic node
of the distinguished diagram.

\subsection{The super L-operator $\mathcal{L}_{C(n)}^{\mathbf{\protect \mu }%
_{n}}$}

The second possible 3-grading for the orthosymplectic $C(n)$\ is associated
to the node $2\mathrm{\delta }_{n-1}$ dual to the coweight $\mathbf{\mu }%
_{n}.$\ It reads in Lie superalgebra language as%
\begin{equation}
osp(2|2n-2)\quad \rightarrow \quad A_{1}\oplus sl(1|n-1)\oplus N_{+}\oplus
N_{-}
\end{equation}%
with
\begin{equation}
N_{\pm }=\frac{n(n-1)}{2}+(n-1)
\end{equation}%
in agreement with the dimensions splitting%
\begin{equation}
2n^{2}+n-2\quad \rightarrow \quad 1+\left( n^{2}-1\right) +N_{+}+N_{-}
\end{equation}%
In this case, the fundamental representation splits as%
\begin{equation}
\left( \mathbf{2|2n-2}\right) \quad \rightarrow \quad \left( \mathbf{1|n-1}%
\right) _{+\frac{1}{2}}\oplus \left( \mathbf{1|n-1}\right) _{-\frac{1}{2}}
\end{equation}%
which is thought of as,%
\begin{equation}
\left( \mathbf{2|2n-2}\right) \quad \rightarrow \quad \left( \mathbf{1|0}%
\right) _{+\frac{1}{2}}\oplus \left( \mathbf{0|n-1}\right) _{+\frac{1}{2}%
}\oplus \left( \mathbf{0|n-1}\right) _{-\frac{1}{2}}\oplus \left( \mathbf{1|0%
}\right) _{-\frac{1}{2}}
\end{equation}%
meaning that we can work in a basis of the form
\begin{equation}
\left \vert 2n\right \rangle \quad \rightarrow \quad \left \vert 0\right
\rangle \oplus \left \vert i\right \rangle \oplus \left \vert \bar{\imath}%
\right \rangle \oplus \left \vert \bar{0}\right \rangle
\end{equation}%
where $1\leq i\leq n-1$ and $\overline{n-1}\leq \bar{\imath}\leq \bar{1}$
with $\bar{\imath}=2n-1-i.$\ We define projectors on these four subspaces as
follows%
\begin{equation}
\varrho =\left \vert 0\right \rangle \left \langle 0\right \vert ,\qquad \Pi
=\sum_{i=1}^{n-1}\left \vert i\right \rangle \left \langle i\right \vert
,\qquad \bar{\Pi}=\sum_{\bar{\imath}=2n-2}^{n}\left \vert \bar{\imath}\right
\rangle \left \langle \bar{\imath}\right \vert ,\qquad \bar{\varrho}=\left
\vert \bar{0}\right \rangle \left \langle \bar{0}\right \vert
\end{equation}%
and therefore the action of the coweight reads as%
\begin{equation}
\mathbf{\mu }_{n}\mathbf{=}\frac{1}{2}\varrho +\frac{1}{2}\Pi -\frac{1}{2}%
\bar{\Pi}-\frac{1}{2}\bar{\varrho}
\end{equation}%
Each of the nilpotents $N_{+}$ and $N_{-}$ split as $%
(n-1)+(n-1)+(n-1)(n-2)/2 $, they are generated by couples $\left( X_{\bar{%
\imath}},X_{\left[ i\bar{j}\right] }\right) $\ and\ $\left( Y^{i},Y^{\left[
\bar{\imath}j\right] }\right) $ where $X_{\bar{\imath}}$ and $Y^{i}$ are
simply realized as%
\begin{equation}
\begin{tabular}{lll}
$X_{i}$ & $=$ & $\left \vert 0\right \rangle \left \langle \bar{\imath}%
\right \vert -\left \vert i\right \rangle \left \langle \bar{0}\right \vert $
\\
$Y^{\bar{\imath}}$ & $=$ & $\left \vert \bar{0}\right \rangle \left \langle
i\right \vert -\left \vert \bar{\imath}\right \rangle \left \langle 0\right
\vert $%
\end{tabular}%
\end{equation}%
while the $X_{\left[ i\bar{j}\right] }$\ and $Y^{\left[ \bar{\imath}j\right]
}$\ are \textrm{anti-}symmetric in $i$ and $j$ and are given by%
\begin{equation}
\begin{tabular}{lll}
$X_{\left[ i\bar{j}\right] }$ & $=$ & $\left \vert i\right \rangle \left
\langle \bar{j}\right \vert -\left \vert j\right \rangle \left \langle \bar{%
\imath}\right \vert $ \\
$Y^{\left[ \bar{\imath}j\right] }$ & $=$ & $\left \vert \bar{\imath}\right
\rangle \left \langle j\right \vert -\left \vert \bar{j}\right \rangle \left
\langle i\right \vert $%
\end{tabular}%
\end{equation}%
We eventually have%
\begin{equation}
\begin{tabular}{lll}
$X$ & $=$ & $b^{i}X_{i}+b^{\left[ i\bar{j}\right] }X_{\left[ i\bar{j}\right]
}$ \\
$Y$ & $=$ & $c_{\bar{\imath}}Y^{\bar{\imath}}+c_{\left[ \bar{\imath}j\right]
}Y^{\left[ \bar{\imath}j\right] }$%
\end{tabular}%
\end{equation}%
which verify%
\begin{equation}
\left[ \mathbf{\mu }_{n}\mathbf{,}X\right] =X\qquad ,\qquad \left[ \mathbf{%
\mu }_{n}\mathbf{,}Y\right] =-Y
\end{equation}%
as well as the nilpotency properties $X^{2}=Y^{2}=0$ leading to $e^{X}=1+X$\
and\ $e^{Y}=1+Y.$ Notice that the oscillators here $\left( b^{i},c_{\bar{%
\imath}}\right) $ are of fermionic nature, while $\left( b^{\left[ i\bar{j}%
\right] },c_{\left[ \bar{\imath}j\right] }\right) $ are bosonic. We further
have%
\begin{equation}
X\Pi =X\varrho =0\qquad ,\qquad \varrho Y=\Pi Y=0
\end{equation}%
and%
\begin{equation}
z^{\mathbf{\mu }_{n}}=z^{\frac{1}{2}}\varrho +z^{\frac{1}{2}}\Pi +z^{-\frac{1%
}{2}}\bar{\Pi}+z^{-\frac{1}{2}}\bar{\varrho}
\end{equation}%
This simplifies the expression of the L-operator as follows%
\begin{equation}
\begin{tabular}{lll}
$\mathcal{L}_{C(n)}^{\mathbf{\mu }_{n}}$ & $=$ & $\left( 1+X\right) \left(
z^{\frac{1}{2}}\varrho +z^{\frac{1}{2}}\Pi +z^{-\frac{1}{2}}\bar{\Pi}+z^{-%
\frac{1}{2}}\bar{\varrho}\right) \left( 1+Y\right) $ \\
& $=$ & $z^{\frac{1}{2}}\varrho +z^{-\frac{1}{2}}\Pi +z^{\frac{1}{2}}\bar{\Pi%
}+z^{-\frac{1}{2}}\bar{\varrho}+z^{-\frac{1}{2}}X\left( \bar{\varrho}+\bar{%
\Pi}\right) +$ \\
&  & $z^{-\frac{1}{2}}\left( \bar{\varrho}+\bar{\Pi}\right) Y+z^{-\frac{1}{2}%
}X\bar{\Pi}Y$%
\end{tabular}%
\end{equation}%
This $\mathcal{L}_{C(n)}^{\mathbf{\mu }_{n}}$ is represented in the
projector basis $\left( \varrho ,\Pi ,\bar{\Pi},\bar{\varrho}\right) $ as
follows%
\begin{equation}
\mathcal{L}_{C(n)}^{\mathbf{\mu }_{n}}=\left(
\begin{array}{cccc}
z^{\frac{1}{2}}-z^{-\frac{1}{2}}b^{i}c_{i} & z^{-\frac{1}{2}}b^{i}c_{ij} &
z^{-\frac{1}{2}}b^{i} & 0 \\
-z^{-\frac{1}{2}}b^{ij}c_{i} & z^{\frac{1}{2}}\mathbf{1}_{n-1}+z^{-\frac{1}{2%
}}\Phi \Psi & z^{-\frac{1}{2}}\Phi & -z^{-\frac{1}{2}}b^{i} \\
z^{-\frac{1}{2}}c_{i} & z^{-\frac{1}{2}}\Psi & z^{-\frac{1}{2}}\mathbf{1}%
_{n-1} & 0 \\
0 & -z^{-\frac{1}{2}}c_{i} & 0 & z^{-\frac{1}{2}}%
\end{array}%
\right)  \label{23}
\end{equation}%
where $\Phi =b^{\left[ i\bar{j}\right] }X_{\left[ i\bar{j}\right] }$ and $%
\Psi =c_{\left[ \bar{\imath}j\right] }Y^{\left[ \bar{\imath}j\right] }$ are
\textrm{anti-}symmetric $(n-1)(n-1)$ matrices given by%
\begin{equation}
\Phi =\left(
\begin{array}{cccc}
0 & b^{1\overline{2}} & \ldots & b^{1\overline{\left( n-1\right) }} \\
-b^{1\overline{2}} & \ddots &  & \vdots \\
\vdots &  & \ddots & b^{(n-2)\overline{\left( n-1\right) }} \\
-b^{1\overline{\left( n-1\right) }} & \ldots & -b^{(n-2)\overline{\left(
n-1\right) }} & 0%
\end{array}%
\right)
\end{equation}%
and%
\begin{equation}
\Psi =\left(
\begin{array}{cccc}
0 & c_{\bar{1}2} & \ldots & c_{\bar{1}\left( n-1\right) } \\
-c_{\bar{1}2} & \ddots &  & \vdots \\
\vdots &  & \ddots & c_{\overline{\left( n-2\right) }\left( n-1\right) } \\
-c_{\bar{1}\left( n-1\right) } & \ldots & -c_{\overline{\left( n-2\right) }%
\left( n-1\right) } & 0%
\end{array}%
\right)  \label{230}
\end{equation}

\section{Super L-operators of $D(m|n)$ type}

\label{sec:6} 
\begin{figure}[tbph]
\begin{center}
\includegraphics[width=14cm]{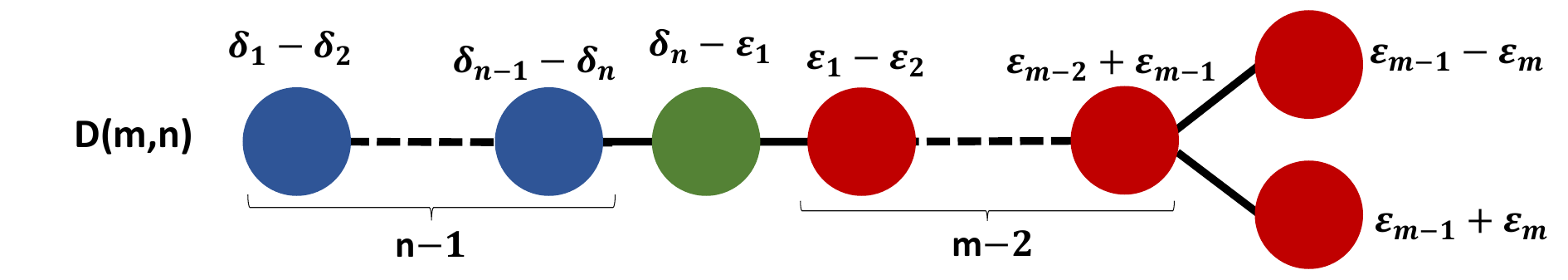}
\end{center}
\par
\vspace{-0.5cm}
\caption{{}Distinguished Dynkin diagram of the $D(m|n)$ superalgebra having
one fermionic simple root in Green color. Here, we have $\protect \varepsilon %
_{i}^{2}=1$ and $\protect \delta _{i}^{2}=-1.$}
\label{D1}
\end{figure}

In this section, we focus on the basic Lie superalgebra of $%
D(m|n)$ type in order to compute its corresponding super L-operators
characterizing $D$-type superspin chains with the internal symmetry%
\begin{equation}
D(m|n)=osp(2m|2n),\qquad m\geq 3,n\geq 1
\end{equation}%
having $2(m+n)^{2}-m+n$ dimensions and the rank $r_{D(m|n)}=m+n$. It is
defined by an even part $D(m|n)_{\bar{0}}$ reading as%
\begin{equation}
D_{m}\oplus C_{n}=so(2m)\oplus sp(2n)
\end{equation}%
and an odd part $D(m|n)_{\bar{1}}$ generated by the bi-fundamental $(2m,2n)$
representation of $D(m|n)_{\bar{0}}.$\ The $D(m|n)$\ superalgebra has
multiple graphical descriptions with $m+n$ nodes represented by graded
simple roots $\left \{ \tilde{\alpha}_{i}\right \} _{1\leq i\leq m+n}$ .
These are generated in terms of $m+n$ fundamental unit weights given by the
bosonic $\left \{ \varepsilon _{a}\right \} _{1\leq a\leq m}$ realising the
roots of $so(2m),$ and the fermionic $\left \{ \mathrm{\delta }_{\text{%
\textsc{a}}}\right \} _{1\leq \text{\textsc{a}}\leq n}$\ realising the roots
of $sp(2n)$; their mixing gives fermionic roots of $D(m|n)_{\bar{1}}$. In
these regards, recall that the super root system $\Phi _{D(m|n)}$ of the Lie
superalgebra $D(m|n)$ has $2(m+n)^{2}-2m\ $roots that split into an even set
$\Phi _{\bar{0}}$ and an odd part $\Phi _{\bar{1}}$ with content \textrm{%
reading as}%
\begin{equation}
\begin{tabular}{l|l|ll}
\ $\Phi _{D(m|n)}$ & \  \  \ roots & \  \  \ number & range of Labels \\
\hline \hline
$\  \  \Phi _{\bar{0}}$ \  \  & $\
\begin{array}{c}
\pm \widetilde{\alpha }_{ab}^{\pm } \\
\pm \widetilde{\beta }_{\text{\textsc{ab}}}^{\pm } \\
\pm 2\mathrm{\delta }_{\text{\textsc{a}}}%
\end{array}%
$ \  \  & $\  \  \
\begin{array}{c}
2m^{2}-2m \\
2n^{2}-2n \\
2n%
\end{array}%
$ \  \  & $\left.
\begin{array}{c}
a\neq b \\
\text{\textsc{a}}\neq \text{\textsc{b}} \\
\end{array}%
\right. $ \\ \hline
$\  \  \Phi _{\bar{1}}$ & $\  \left.
\begin{array}{c}
\widetilde{\gamma }_{a\text{\textsc{b}}}^{\pm } \\
-\widetilde{\gamma }_{a\text{\textsc{b}}}^{\pm } \\
\text{ \  \ }%
\end{array}%
\right. $ & $\  \  \  \  \  \  \left.
\begin{array}{c}
2mn \\
2mn \\
\end{array}%
\right. $ & $\left.
\begin{array}{c}
a=1,...,m \\
\text{\textsc{a}}=1,...,n \\
\end{array}%
\right. $ \\ \hline \hline
\end{tabular}%
\end{equation}%
where we have set%
\begin{equation}
\widetilde{\alpha }_{ab}^{\pm }=\varepsilon _{a}\pm \varepsilon _{b},\qquad
\widetilde{\beta }_{\text{\textsc{ab}}}^{\pm }=\mathrm{\delta }_{\text{%
\textsc{a}}}\pm \mathrm{\delta }_{\text{\textsc{b}}},\qquad \widetilde{%
\gamma }_{a\text{\textsc{b}}}^{\pm }=\varepsilon _{a}\pm \mathrm{\delta }_{%
\text{\textsc{b}}}
\end{equation}%
A remarkable simple root basis generating the super root system $\Phi
_{D(m|n)}$ is given by the distinguished basis $(\mathrm{\beta }_{\text{%
\textsc{a}}},\mathrm{\gamma ,}\alpha _{a})$ having one fermionic root $%
\mathrm{\gamma }$ as given here below
\begin{equation}
\begin{tabular}{lllll}
$\mathrm{\beta }_{\text{\textsc{a}}}$ & $=$ & $\mathrm{\delta }_{\text{%
\textsc{a}}}-\mathrm{\delta }_{\text{\textsc{a+1}}}$ & $,$ & $\text{\textsc{a%
}}=1,...,n-1$ \\
$\alpha _{a}$ & $=$ & $\varepsilon _{a}-\varepsilon _{a+1}$ & $,$ & $%
a=1,...,m-1$ \\
$\alpha _{m}$ & $=$ & $\varepsilon _{m-1}+\varepsilon _{m}$ &  &  \\
$\mathrm{\gamma }$ & $=$ & $\mathrm{\delta }_{n}-\varepsilon _{1}$ &  &
\end{tabular}%
\end{equation}%
This simple root basis can be collectively denoted shortly as $\tilde{\alpha}%
_{i}=(\widetilde{\beta }_{\text{\textsc{a}}},\widetilde{\gamma }_{n}\mathrm{,%
}\widetilde{\alpha }_{a})$ with super label $i=1,...,m+n.$ Notice that this
distinguished basis is characterised by the ordering of the set $\left \{
\mathrm{\delta }_{\text{\textsc{a}}},\varepsilon _{a}\right \} $ of the
fundamental unit weight vectors as follows%
\begin{equation}
\mathrm{\delta }_{1},\quad \mathrm{\delta }_{2},\quad ...\quad \mathrm{%
\delta }_{n-1},\quad \mathrm{\delta }_{n};\quad \varepsilon _{1},\quad
\varepsilon _{2},\quad ...\quad \varepsilon _{m-1},\quad \varepsilon _{m}
\label{ord}
\end{equation}%
leading in turn to an ordering of the set of graded simple root as $(%
\widetilde{\beta }_{\text{\textsc{a}}},\widetilde{\gamma }_{n}\mathrm{,}%
\widetilde{\alpha }_{a}),$ and consequently to the Distinguished Dynkin
diagram depicted in \textbf{Figure \ref{D1}} where the graded simple roots
are also reported.
Notice that the basis in eq(\ref{ord}) gives a very particular ordering of
the set $\left \{ \mathrm{\delta }_{\text{\textsc{a}}},\varepsilon
_{a}\right \} $ where all the $\mathrm{\delta }_{\text{\textsc{a}}}$'s are
put on the left and all the $\varepsilon _{a}$'s are on the right. In the
general case where the $\mathrm{\delta }_{\text{\textsc{a}}}$'s and the $%
\varepsilon _{a}$'s are mixed, there are
\begin{equation}
N_{D(m|n)}=\frac{\left( n+m\right) !}{n!\times m!}
\end{equation}%
possibilities of orderings of the type (\ref{ord}). This variety of
orderings indicates that generally speaking the Lie superalgebra $D(m|n)$
has $N_{D(m|n)}$ possible super Dynkin diagrams
\begin{equation}
\mathfrak{D}[D(m|n)]^{\left( \mathrm{\kappa }\right) }\qquad ,\qquad \mathrm{%
\kappa }=1,...,N_{D(m|n)}
\end{equation}%
and eventually $N_{D(m|n)}$\ of varieties of superspin chains of D-type. For
the example of $D(3|1)=osp(6|2)$, we have $\frac{4!}{3!\times 1!}=4$
possible Dynkin diagrams as in Table 1 of \cite{str}.\ For the present
study, we use the 3-gradings in Table eq(\ref{219}$)$\ in order to generate
two types of Lax operators for the distinguished superspin chain with $%
D(m|n) $\ symmetry.\ The first one is referred to as the spinorial
L-operator $\mathcal{L}_{D_{m|n}}^{\mathbf{\mu }_{m+n}}$\ because it is
linked to the (co)spinorial nodes of the super diagram \textbf{\ref{D1}}.
The second one is labeled as $\mathcal{L}_{D_{m|n}}^{\mathbf{\mu }_{n+1}}$\
since it concerns the coweight $\mathbf{\mu }_{n+1}$ associated to $%
\varepsilon _{1}-\varepsilon _{2}$\ in \textbf{\ref{D1}}\textrm{\textbf{.}}

\subsection{The super L-operator $\mathcal{L}_{D_{m|n}}^{\mathbf{\protect \mu
}_{m+n}}$}

Due to the $Z_{2}$\ automorphism symmetry of the distinguished super Dynkin
diagram of the \textbf{Figure \ref{D1}} which permutes the spinor and
cospinor roots $\widetilde{\alpha }_{m+n}$\ and $\widetilde{\alpha }_{m+n-1}$%
, we can deduce that the coweights $\mathbf{\mu }_{m+n}$\ and $\mathbf{\mu }%
_{m+n-1}$\ act in the same way on the Lie superalgebra $osp(2m|2n)$\ and
therefore we have similar super L-operators. In fact, the 3-grading obtained
by the cutting of one of the these nodes in the distinguished super Dynkin
diagram is the same. We have

\begin{equation}
osp(2m|2n)\quad \longrightarrow \quad gl\left( 1\right) \oplus sl(m|n)\oplus
N_{+}\oplus N_{-}  \label{spsl}
\end{equation}%
with the super special linear $sl(m|n)$ as a sub- superalgebra. For this
breaking pattern, we have the dimensions%
\begin{equation}
\begin{tabular}{|c|c|c|c|c|c|}
\hline
{\small algebra} & ${\small osp}_{{\small 2m|2n}}$ & ${\small so}_{2}$ & $%
{\small sl(m|n)}$ & ${\small N}_{+}$ & ${\small N}_{-}$ \\ \hline
$\dim $ & $\left.
\begin{array}{c}
\text{ \  \  \ } \\
{\small 2(m+n)}^{2}{\small -m+n} \\
\text{ \  \  \ }%
\end{array}%
\right. $ & ${\small 1}$ & ${\small (m+n)}^{2}{\small -1}$ & $\frac{{\small %
(m+n)}^{2}+n{\small -m}}{2}$ & $\frac{{\small (m+n)}^{2}+n{\small -m}}{2}$
\\ \hline
\end{tabular}
\label{SOO}
\end{equation}%
The fundamental representation $\left( \mathbf{2m|2n}\right) $ of the
superalgebra $osp(2m|2n)$ splits in terms of the fundamental $\left( \mathbf{%
m|n}\right) _{\pm q}$ representations of $sl(m|n)$ as follows%
\begin{equation}
\left( \mathbf{2m|2n}\right) =\left( \mathbf{m|n}\right) _{+\frac{1}{2}%
}\oplus \left( \mathbf{\bar{m}|\bar{n}}\right) _{-\frac{1}{2}}  \label{15}
\end{equation}%
This decomposition means that the super L-operator $\mathcal{L}_{D_{m|n}}^{%
\mathbf{\mu }_{m+n}}$ (and equivalently $\mathcal{L}_{D_{m|n}}^{\mathbf{\mu }%
_{m+n-1}}$) will be represented by a $2(m+n)\times 2(m+n)$ matrix
\begin{equation}
\left(
\begin{array}{cc}
L_{\text{\textsc{ij}}} & L_{\text{\textsc{i\={j}}}} \\
L_{\text{\textsc{\={\i}j}}} & L_{\text{\textsc{\={\i}\={j}}}}%
\end{array}%
\right)
\end{equation}%
with basis vector of the form
\begin{equation}
\begin{tabular}{llll}
$\left \vert \text{\textsc{i}}\right \rangle $ & $\equiv \left \vert i\right
\rangle \oplus \left \vert \alpha \right \rangle $ & \quad for\quad & $%
\left( \mathbf{m|n}\right) _{+\frac{1}{2}}$ \\
$\left \vert \text{\textsc{\={\i}}}\right \rangle $ & $\equiv \left \vert
\bar{\imath}\right \rangle \oplus \left \vert \bar{\alpha}\right \rangle $ &
\quad for\quad & $\left( \mathbf{\bar{m}|\bar{n}}\right) _{-\frac{1}{2}}$%
\end{tabular}
\label{16}
\end{equation}%
In this basis, the subscripts $\pm 1/2$ are charges of $gl(1)$. The $\left
\vert i\right \rangle $\ and $\left \vert \bar{\imath}\right \rangle $\
generate the (anti-)fundamental representations $\mathbf{m}$\ and $\mathbf{%
\bar{m}}$ of $sl(m);$ while the $\left \vert \alpha \right \rangle $ and $%
\left \vert \bar{\alpha}\right \rangle $ generate the (anti-)fundamental
representations $\mathbf{n}$\ and $\mathbf{\bar{n}}$ of $sl(n)$. Notice that
by ordering the basis vectors of $osp(2m|2n)$ like $\left( \left \vert
i\right \rangle ,\left \vert \alpha \right \rangle ,\left \vert \bar{\imath}%
\right \rangle ,\left \vert \bar{\alpha}\right \rangle \right) ,$ the labels
\textrm{take} the values%
\begin{equation}
\begin{tabular}{lll}
$1\leq i\leq m$ & $,\qquad $ & $n+m+1\leq \bar{\imath}\leq n+2m$ \\
$1\leq \alpha \leq n$ & $,\qquad $ & $2m+n+1\leq \bar{\alpha}\leq 2n+2m$%
\end{tabular}%
\end{equation}%
In terms of the super labels \textsc{i} and \textsc{\={\i}} introduced in (%
\ref{16}), we can rewrite these intervals in a short way like%
\begin{equation}
1\leq \text{\textsc{i}}\leq m+n,\qquad n+m+1\leq \text{\textsc{\={\i}}}\leq
2n+2m
\end{equation}%
Using these super labels, we can construct the operators \textrm{involved in}
the expression of the super Lax operators $\mathcal{L}^{\mathbf{\mu }%
_{m+n}}=e^{X}z^{\mathbf{\mu }_{m+n}}e^{Y}$ associated with the breaking
pattern (\ref{spsl})\textrm{\ of }the \textrm{distinguished} diagram of the
\textbf{Figure \ref{D1}}.\newline
First, from eq(\ref{15}) we learn that the action of the coweight $\mathbf{%
\mu }_{m+n}$ is given by%
\begin{equation}
\mathbf{\mu }_{m+n}\mathbf{=}\frac{1}{2}\sum_{\text{\textsc{i}}%
=1}^{m+n}\left \vert \text{\textsc{i}}\right \rangle \left \langle \text{%
\textsc{i}}\right \vert -\frac{1}{2}\sum_{\text{\textsc{\={\i}}}%
=m+n+1}^{2m+2n}\left \vert \text{\textsc{\={\i}}}\right \rangle \left
\langle \text{\textsc{\={\i}}}\right \vert  \label{pro}
\end{equation}%
which, by setting $\mathbf{\Pi }=\sum_{\text{\textsc{i}}=1}^{m+n}\left \vert
\text{\textsc{i}}\right \rangle \left \langle \text{\textsc{i}}\right \vert $
and $\bar{\Pi}=\sum_{\text{\textsc{\={\i}}}=m+n+1}^{2m+2n}\left \vert \text{%
\textsc{\={\i}}}\right \rangle \left \langle \text{\textsc{\={\i}}}\right
\vert $ reads also as $\mathbf{\mu }_{m+n}\mathbf{=}\frac{1}{2}\Pi -\frac{1}{%
2}\bar{\Pi}.$ The $\mathbf{\Pi }$ and $\bar{\Pi}$\ are projectors of the
fundamental representations of $sl(m|n)$. In terms of $sl(m)\oplus sl(n)$
vector basis, the $\mathbf{\mu }_{m+n}$ splits as follows%
\begin{equation}
\mathbf{\mu }_{m+n}\mathbf{=}\frac{1}{2}\left( \sum_{i=1}^{m}\left \vert
i\right \rangle \left \langle i\right \vert +\sum_{\alpha =m+1}^{m+n}\left
\vert \alpha \right \rangle \left \langle \alpha \right \vert \right) -\frac{%
1}{2}\left( \sum_{\bar{\imath}=m+n+1}^{2m+n}\left \vert \bar{\imath}\right
\rangle \left \langle \bar{\imath}\right \vert +\sum_{\bar{\alpha}%
=2m+n+1}^{2m+2n}\left \vert \bar{\alpha}\right \rangle \left \langle \bar{%
\alpha}\right \vert \right)  \label{11}
\end{equation}%
Second, the matrix operators $X$ and $Y$ in the expression $e^{X}z^{\mathbf{%
\mu }_{m+n}}e^{Y}$ belong respectively to the nilpotents $N_{+}$ and $N_{-}$%
; they can be expanded in terms of representations of $sl_{m}\oplus sl\left(
n\right) .$ This feature follows from the decomposition of $\dim N_{\pm }$ (%
\ref{SOO})
\begin{equation}
\dim N_{\pm }=\frac{m(m-1)}{2}+\frac{n(n+1)}{2}+mn
\end{equation}%
involving the antisymmetric representation of $sl(m)$, the symmetric
representation of $sl(n)$ and the bi-fundamental representation. Thus, an
explicit realization of $X$ is given by using the generators $(X_{\left[ i%
\bar{j}\right] },X_{{\small (\alpha \bar{\beta})}},\mathcal{X}_{i\bar{\alpha}%
})$ and the graded Darboux coordinates $(b^{\left[ i\bar{j}\right] },\mathrm{%
f}^{{\small (\alpha \bar{\beta})}},\mathrm{\beta }^{i\bar{\alpha}})$ as
follows%
\begin{equation}
X=b^{\left[ i\bar{j}\right] }X_{\left[ i\bar{j}\right] }+\mathrm{f}^{{\small %
(\alpha \bar{\beta})}}X_{\left( \alpha \bar{\beta}\right) }+\mathrm{\beta }%
^{i\bar{\alpha}}\mathcal{X}_{\left[i\bar{\alpha}\right]}  \label{12}
\end{equation}%
with%
\begin{equation}
\begin{tabular}{lll}
$X_{\left[ i\bar{j}\right] }$ & $=$ & $\left \vert i\right \rangle \left
\langle \bar{j}\right \vert -\left \vert j\right \rangle \left \langle \bar{%
\imath}\right \vert $ \\
$X_{\left( \alpha \bar{\beta}\right) }$ & $=$ & $\left \vert \alpha \right
\rangle \left \langle \bar{\beta}\right \vert +\left \vert \beta \right
\rangle \left \langle \bar{\alpha}\right \vert $ \\
$\mathcal{X}_{\left[i\bar{\alpha}\right]}$ & $=$ & $\left \vert i\right \rangle \left
\langle \bar{\alpha}\right \vert - \left \vert \alpha \right \rangle \left
\langle \bar{i}\right \vert 

$
\end{tabular}%
\end{equation}%
verifying the nilpotency property $X^{2}=0$ and indicating that $e^{x}=I+X$.
Similarly for the Y operator belonging to the nilpotent $N_{-}$, we have $%
Y^{2}=0$ with the expansion%
\begin{equation}
Y=c_{\left[ \bar{j}i\right] }Y^{\left[ \bar{j}i\right] }+\mathrm{g}_{{\small %
(\bar{\beta}\alpha )}}Y^{{\small (\alpha \bar{\beta})}}+\mathrm{\gamma }_{%
\bar{\alpha}i}\mathcal{Y}^{\left[\bar{\alpha}i\right]}  \label{13}
\end{equation}%
where $(c_{\left[ \bar{j}i\right] },\mathrm{g}_{\left( \bar{\beta}\alpha
\right) },\mathrm{\gamma }_{\bar{\alpha}i})$ are graded Darboux coordinates
that are conjugate to ($b^{\left[ i\bar{j}\right] },\mathrm{f}^{\left(
\alpha \bar{\beta}\right) },\mathrm{\beta }^{i\bar{\alpha}}$) and where
\begin{equation}
\begin{tabular}{lll}
$Y^{\left[ \bar{j}i\right] }$ & $=$ & $\left \vert \bar{j}\right \rangle
\left \langle i\right \vert -\left \vert \bar{\imath}\right \rangle \left
\langle j\right \vert $ \\
$Y^{\left( \bar{\beta}\alpha \right) }$ & $=$ & $\left \vert \bar{\beta}%
\right \rangle \left \langle \alpha \right \vert +\left \vert \bar{\alpha}%
\right \rangle \left \langle \beta \right \vert $ \\
$\mathcal{Y}^{\left[\bar{\alpha}i\right]}$ & $=$ & $\left \vert \bar{\alpha}\right
\rangle \left \langle i\right \vert - \left \vert \bar{i}\right
\rangle \left \langle \alpha \right \vert
$%
\end{tabular}%
\end{equation}%
In total, we have $2[m(m-1)/2+n(n+1)/2]$\ bosonic oscillators given
by $(b^{\left[ ij\right] },c_{\left[ \bar{j}i\right] }),$ $(\mathrm{f}%
^{\left( \alpha \bar{\beta}\right) },\mathrm{g}_{\left( \bar{\beta}\alpha
\right) }),$ and $2mn$\ fermionic ones given by $(\mathrm{\beta }^{i\bar{%
\alpha}},\mathrm{\gamma }_{\bar{\alpha}i}).$ Using the properties $%
X^{2}=Y^{2}=0$, the super Lax operator reads as follows%
\begin{equation}
\mathcal{L}_{D_{m|n}}^{\mathbf{\mu }_{m+n}}=z^{\mathbf{\mu }_{m+n}}+z^{%
\mathbf{\mu }_{m+n}}Y+Xz^{\mathbf{\mu }_{m+n}}+Xz^{\mathbf{\mu }_{m+n}}Y
\end{equation}%
Using (\ref{pro}), we \textrm{obtain} interesting properties useful for the
calculation of $\mathcal{L}_{D_{m|n}}^{\mathbf{\mu }_{m+n}}$. The charge
operator $z^{\mathbf{\mu }}$ reads as $z^{\frac{1}{2}\Pi -\frac{1}{2}\bar{\Pi%
}}$; and the nilpotent operators $X$ and $Y$ obey the properties
\begin{eqnarray}
\mathbf{\Pi }X &=&X,\qquad X\mathbf{\Pi }=0,\qquad \mathbf{\bar{\Pi}}%
Y=Y,\qquad Y\mathbf{\bar{\Pi}}=0 \\
X\mathbf{\bar{\Pi}} &=&X,\qquad \mathbf{\bar{\Pi}}X=0,\qquad Y\mathbf{\Pi }%
=Y,\qquad Y\mathbf{\Pi }=0
\end{eqnarray}%
The super L-operator is calculated by substituting in (\ref{lax}) with (\ref%
{11}-\ref{13}) and using $Xz^{\mathbf{\mu }_{m+n}}=z^{-\frac{1}{2}}X$ and $%
z^{\mathbf{\mu }_{m+n}}Y=z^{-\frac{1}{2}}Y$; we have%
\begin{equation*}
\begin{tabular}{lll}
$\mathcal{L}_{D_{m|n}}^{\mathbf{\mu }_{m+n}}$ & $=$ & $z^{\mathbf{\mu }%
_{m+n}}+z^{-\frac{1}{2}}X+z^{-\frac{1}{2}}Y$ \\
&  & $+z^{-\frac{1}{2}}(\Phi \Psi +\Lambda \Gamma +\mathrm{f}^{\left( \alpha
\bar{\beta}\right) }\mathrm{\gamma }_{\bar{\alpha}i}X_{\left( \alpha \bar{%
\beta}\right) }\mathcal{Y}^{\left[\bar{\alpha}i\right]}+\mathrm{\beta }^{i\bar{\alpha}}%
\mathrm{g}_{\left( \bar{\beta}\alpha \right) }\mathcal{X}_{\left[i\bar{\alpha}\right]}Y^{\left( \bar{\beta}\alpha \right) })$ \\
&  & $+z^{-\frac{1}{2}}\mathrm{\beta }^{i\bar{\alpha}}\mathrm{\gamma }_{\bar{%
\alpha}i}\mathcal{X}_{\left[i\bar{\alpha}\right]}\mathcal{Y}^{\left[\bar{\alpha}i\right]}$%
\end{tabular}%
\end{equation*}%
where we have set%
\begin{equation}
\begin{tabular}{lll}
$\Phi =b^{\left[ i\bar{j}\right] }X_{\left[ i\bar{j}\right] }$ & $\qquad
,\qquad $ & $\Psi =c_{\left[ \bar{j}i\right] }Y^{\left[ \bar{j}i\right] }$
\\
$\Lambda =\mathrm{f}^{\left( \alpha \bar{\beta}\right) }X_{\left( \alpha
\bar{\beta}\right) }$ & $\qquad ,\qquad $ & $\Gamma =\mathrm{g}_{\left( \bar{%
\beta}\alpha \right) }Y^{\left( \bar{\beta}\alpha \right) }$%
\end{tabular}
\label{17}
\end{equation}%
The matrix form after multiplying with the overall factor $z^{\frac{1}{2}}$
is given in the basis ($\left \vert i\right \rangle ,\left \vert \alpha
\right \rangle ,\left \vert \bar{\alpha}\right \rangle ,\left \vert \bar{%
\imath}\right \rangle $) by%
\begin{equation}
    \mathcal{L}_{D(m|n)}^{\mathbf{\mu }_{m+n}}=\left(\begin{array}{cccc}z\delta _{j}^{i}+\Phi \Psi +\mathrm{\beta }^{i\bar{\alpha}}\mathrm{\gamma }_{\bar{\alpha}i} & \mathrm{\beta }^{i\bar{\alpha}}\mathrm{g}_{\left( \bar{\beta}\alpha \right) } & \mathrm{\beta }^{i\bar{\alpha}} & \Phi \\
    \mathrm{f}^{\left( \alpha \bar{\beta}\right) }\mathrm{\gamma }_{\bar{\alpha}i} & z\delta _{\beta }^{\alpha }+\Lambda \Gamma & \Lambda & -\mathrm{\beta }^{i\bar{\alpha}} \\
    \mathrm{\gamma }_{\bar{\alpha}i} & \Gamma & z\delta _{\bar{\beta}}^{\bar{\alpha}}& 0 \\
    \Psi & -\mathrm{\gamma }_{\bar{\alpha}i} & 0 & z\delta _{\bar{j}}^{\bar{\imath}}
    \end{array}
    \right) \label{25}
\end{equation}

The $\Phi ,\Psi $ in (\ref{17}) are anti-symmetric $(m\times m)$ matrices
while $\Lambda $ and $\Gamma $ are $(n\times n)$ symmetric matrices reading
explicitly as%
\begin{equation}
\Phi =\left(
\begin{array}{ccc}
0 & \ldots & b^{1\overline{m}} \\
\vdots & \ddots & \vdots \\
b^{m\bar{1}} & \ldots & 0%
\end{array}%
\right) \qquad ;\qquad \Psi =\left(
\begin{array}{ccc}
0 & \ldots & c_{\bar{1}m} \\
\vdots & \ddots & \vdots \\
c_{\overline{m}1} & \ldots & 0%
\end{array}%
\right)
\end{equation}%
and%
\begin{equation}
\Lambda =\left(
\begin{array}{ccc}
\mathrm{f}^{1\bar{1}} & \ldots & \mathrm{f}^{1\overline{n}} \\
\vdots & \ddots & \vdots \\
\mathrm{f}^{n\bar{1}} & \ldots & \mathrm{f}^{n\overline{n}}%
\end{array}%
\right) \qquad ;\qquad \Gamma =\left(
\begin{array}{ccc}
\mathrm{g}_{\bar{1}1} & \ldots & \mathrm{g}_{\bar{1}n} \\
\vdots & \ddots & \vdots \\
\mathrm{g}_{\overline{n}1} & \ldots & \mathrm{g}_{\overline{n}n}%
\end{array}%
\right)  \label{250}
\end{equation}

\subsection{The super L-operator $\mathcal{L}_{D_{m|n}}^{\mathbf{\protect \mu
}_{n+1}}$}

Now, we move to the investigation of the super Lax operator $\mathcal{L}%
_{D_{m|n}}^{\mathbf{\mu }_{n+1}}$\ associated with the second possible
3-grading of the Lie superalgebra $osp(2m|2n).$\ On the level of the
distinguished Dynkin diagram, this decomposition is associated to the node\
of the simple root $\widetilde{\alpha }_{1}=\varepsilon _{1}-\varepsilon
_{2} $\ which leads to%
\begin{equation}
osp(2m|2n)\quad \rightarrow \quad N_{+}\oplus \boldsymbol{l}_{\mathbf{\mu }%
}\oplus N_{-}  \label{bpt}
\end{equation}%
with
\begin{equation*}
\boldsymbol{l}_{\mathbf{\mu }}=so(2)\oplus osp(2m-2|2n)
\end{equation*}%
and the \textrm{nilpotents' dimensions} given by
\begin{equation}
\begin{tabular}{lll}
$\dim \boldsymbol{l}_{\mathbf{\mu }}$ & $=$ & ${\small 2(m-1+n)}^{2}{\small %
-m+n+1}$ \\
$\dim N_{\pm }$ & $=$ & $2\left( m-1\right) +2n$%
\end{tabular}%
\end{equation}%
Under this breaking, the fundamental representation $\mathbf{2m|2n}$ of the
orthosymplectic $osp(2m|2n)$ splits in terms of representations of $%
so(2)\oplus osp(2m-2|2n)$ as follows%
\begin{equation}
\begin{tabular}{lll}
$\left( \mathbf{2m|2n}\right) $ & $\quad \rightarrow \quad $ & $\left(
\mathbf{2m-2|2n}\right) _{0}\oplus \left( \mathbf{2|}0\right) _{0}$ \\
$\left( \mathbf{2|0}\right) $ & $\quad \rightarrow \quad $ & $\left( \mathbf{%
1|}0\right) _{+}\oplus \left( \mathbf{1|0}\right) _{-}$%
\end{tabular}
\label{ddec}
\end{equation}%
where we have used the reducibility of $so(2)$ to split the representation $%
\mathbf{2}$ like $\mathbf{1}_{+}\oplus \mathbf{1}_{-}$. By labeling $\left(
\mathbf{2m|2n}\right) $ by the ket vector $\left \vert \text{\textsc{a}}%
\right \rangle $, the decomposition (\ref{ddec}) read in terms of low
dimensional ket vectors as
\begin{subequations}
\begin{equation}
\left \vert \text{\textsc{a}}\right \rangle \quad \rightarrow \quad \left
\vert +\right \rangle \oplus \left \vert A\right \rangle \oplus \left \vert
-\right \rangle  \label{B2}
\end{equation}%
with $\left \vert A\right \rangle =\left \vert i\right \rangle \oplus \left
\vert \alpha \right \rangle $ and $i=1,...2m-2,$ $\alpha =1,...,2n.$ Using
the projectors
\end{subequations}
\begin{equation}
\begin{tabular}{lll}
$\Pi _{+}$ & $=$ & $\left \vert +\right \rangle \left \langle +\right \vert $
\\
$\Pi _{-}$ & $=$ & $\left \vert -\right \rangle \left \langle -\right \vert $%
\end{tabular}%
\end{equation}%
and,%
\begin{equation}
\Pi _{0}=\sum \limits_{A\text{=1}}^{2m+2n-2}\left \vert A\right \rangle
\left \langle A\right \vert =\sum \limits_{i=1}^{2m-2}\left \vert i\right
\rangle \left \langle i\right \vert +\sum \limits_{i=1}^{2n}\left \vert
\alpha \right \rangle \left \langle \alpha \right \vert
\end{equation}%
the action of the coweight reads as follows
\begin{equation}
\mathbf{\mu }_{n+1}=\Pi _{+}+q\Pi _{0}-\Pi _{-}
\end{equation}%
with $q=0$. Similarly, the $2\left( m+n-1\right) $\ generators of the
nilpotent superalgebras $N_{\pm }$ expand like%
\begin{equation}
\begin{tabular}{llllll}
$X$ & $=$ & $\mathcal{B}^{A}\boldsymbol{X}_{A}$ & $=$ & $b^{i}X_{i}+\mathrm{%
\beta }^{\alpha }\mathcal{X}_{\alpha }$ & $\in N_{+}$ \\
$Y$ & $=$ & $\mathcal{C}_{A}\boldsymbol{Y}^{A}$ & $=$ & $c_{i}Y^{i}+\mathrm{%
\gamma }_{\alpha }\mathcal{Y}^{\alpha }$ & $\in N_{-}$%
\end{tabular}
\label{mmm}
\end{equation}%
where the $(b^{i},c_{i})$\ are bosonic Darboux coordinates and $(\mathrm{%
\beta }^{\alpha },\mathrm{\gamma }_{\alpha })$\ are fermionic homologue. The
realisation of the generators of $N_{\pm }$ is given by%
\begin{equation}
\begin{tabular}{lll}
$\boldsymbol{X}_{A}$ & $=$ & $\left \vert \mathbf{+}\right \rangle \left
\langle A\right \vert -\left \vert A\right \rangle \left \langle -\right
\vert $ \\
$\boldsymbol{Y}^{A}$ & $=$ & $\left \vert A\right \rangle \left \langle
\mathbf{+}\right \vert -\left \vert -\right \rangle \left \langle A\right
\vert $%
\end{tabular}%
\end{equation}%
they split like%
\begin{equation}
\begin{tabular}{lll}
$X_{i}$ & $=$ & $\left \vert \mathbf{+}\right \rangle \left \langle i\right
\vert -\left \vert i\right \rangle \left \langle -\right \vert $ \\
$Y^{i}$ & $=$ & $\left \vert i\right \rangle \left \langle \mathbf{+}\right
\vert -\left \vert -\right \rangle \left \langle i\right \vert $%
\end{tabular}%
\qquad ,\qquad
\begin{tabular}{lll}
$\mathcal{X}_{\alpha }$ & $=$ & $\left \vert \mathbf{+}\right \rangle \left
\langle \alpha \right \vert -\left \vert \alpha \right \rangle \left \langle
-\right \vert $ \\
$\mathcal{Y}^{\alpha }$ & $=$ & $\left \vert \alpha \right \rangle \left
\langle \mathbf{+}\right \vert -\left \vert -\right \rangle \left \langle
\alpha \right \vert $%
\end{tabular}%
\end{equation}%
Using these expression, we compute the powers of $X$ and $Y;$ the non
vanishing ones are given by%
\begin{equation}
\begin{tabular}{lll}
$X^{2}$ & $=$ & $-\mathcal{B}^{2}\left \vert \mathbf{+}\right \rangle \left
\langle -\right \vert $ \\
$Y^{2}$ & $=$ & $-\mathcal{C}^{2}\left \vert -\right \rangle \left \langle
\mathbf{+}\right \vert $%
\end{tabular}%
\qquad ,\qquad
\begin{tabular}{lll}
$\mathcal{B}^{2}$ & $=$ & $\left( \mathbf{b}^{2}+\mathbf{\beta }^{2}\right)
\ $ \\
$\mathcal{C}^{2}$ & $=$ & $\left( \mathbf{c}^{2}+\mathbf{\gamma }^{2}\right)
\ $%
\end{tabular}%
\end{equation}%
with%
\begin{equation}
\begin{tabular}{lllllll}
$\mathbf{b}^{2}$ & $=$ & $b^{i}\delta _{ij}b^{j}$ & $\qquad ,\qquad $ & $%
\mathbf{\beta }^{2}$ & $=$ & $\mathrm{\beta }^{\alpha }\delta _{\alpha \beta
}\mathrm{\beta }^{\beta }$ \\
$\mathbf{c}^{2}$ & $=$ & $c_{i}\delta ^{ij}c_{j}$ & $\qquad ,\qquad $ & $%
\mathbf{\gamma }^{2}$ & $=$ & $\mathrm{\gamma }_{\alpha }\delta ^{\alpha
\beta }\mathrm{\gamma }_{\beta }$%
\end{tabular}%
\end{equation}%
Substituting, the super L-operator $\mathcal{L}_{D_{m|n}}^{\mathbf{\mu }%
_{n+1}}=e^{X}z^{\mathbf{\mu }_{n+1}}e^{Y}$ expands as follows%
\begin{equation}
\mathcal{L}_{D_{m|n}}^{\mathbf{\mu }_{n+1}}=(1+X+\frac{X^{2}}{2})\left( z\Pi
_{+}+\Pi _{0}+z^{-1}\Pi _{-}\right) (1+Y+\frac{Y^{2}}{2})
\end{equation}%
In the basis (\ref{B2}) and in terms of bosonic $(b^{i},c_{i})$ and
fermioinc $(\mathrm{\beta }^{\alpha },\mathrm{\gamma }_{\alpha })$
oscillators, we have%
\begin{equation}
\mathcal{L}_{D_{m|n}}^{\mathbf{\mu }_{n+1}}=\left(
\begin{array}{llll}
z^{2}{\small +z(}b^{i}c_{i}+\beta ^{\alpha }\gamma _{\alpha }{\small )+}%
\frac{(\mathbf{b}^{2}\mathbf{+\beta }^{2})(\mathbf{c}^{2}\mathbf{+\gamma }%
^{2})}{4} & zb^{i}{\small +}c_{i}\frac{\mathbf{b}^{2}\mathbf{+\beta }^{2}}{2}
& z\mathrm{\beta }^{\alpha }{\small +}\frac{\left( \mathbf{b}^{2}\mathbf{%
+\beta }^{2}\right) }{2}\mathrm{\gamma }_{\alpha } & \frac{{\small -}(%
\mathbf{b}^{2}\mathbf{+\beta }^{2})}{2} \\
zc_{i}+\frac{(\mathbf{c}^{2}\mathbf{+\gamma }^{2})}{2}b^{i} & z\delta
_{j}^{i}+b^{i}c_{j} & b_{\beta }^{i}\mathrm{\gamma } & -b^{i} \\
z\mathrm{\gamma }_{\alpha }+\frac{(\mathbf{c}^{2}\mathbf{+\gamma }^{2})}{2}%
\mathrm{\beta }^{\alpha } & \mathrm{\beta }^{\alpha }c_{j} & z\delta _{\beta
}^{\alpha }+\mathrm{\beta }^{\alpha }\gamma _{\beta } & -\mathrm{\beta }%
^{\alpha } \\
-\frac{(\mathbf{c}^{2}\mathbf{+\gamma }^{2})}{2} & -c_{i} & -\mathrm{\gamma }%
^{\alpha } & 1%
\end{array}%
\right)  \label{24}
\end{equation}%
where we have multiplied by \textrm{an overall} factor $z$. This matrix has
a very similar structure to (\ref{21}), the only difference concerns the
size of the block of the subspace $\left \vert i\right \rangle $ which is of
$2m-2$ dimensions here.

\section{Conclusion and comments}

\label{sec:7}

The present investigation is an extension of the results of the 4D CS/
Integrability correspondence formulated in \cite{11,1A}, and further
complemented in \cite{costello}. In the latter, XXX spin chains were linked
to a construction of line defects in four dimensional Chern-Simons theory,
and L-operators solving the RLL equation were interpreted as the parallel
transport on the phase space of magnetic 't Hooft line defects. This
correspondence yields a simple and direct formula for the computation of
minuscule L-operators based on Levi decompositions of the bosonic symmetry
algebra, which are in turns directly deduced by cutting minuscule nodes from
the associated Dynkin diagram. This general formula allowed to explicitly
realize oscillator Lax operators for the spin chains with bosonic $ABCDE$
symmetries. Some of these solutions are new to the spin chain literature
while the others perfectly agree with the results obtained from Yangian
based techniques.\newline
The generalization of this correspondence to the super case was initially
treated in \cite{slmn} for the case of $sl(m|n)$ superspin chains. In
analogy to the aforementioned bosonic construction, the oscillator
realizations of super Lax operators solving the RLL equation for a superspin
chain are deduced from\ special decompositions of the Lie superalgebra.
Following this rationale, we constructed in this paper a list of Lax
operators for superspin chains with internal symmetries given by the $ABCD$
Lie superalgebras.\ In this regard, notice that these solutions are obtained
for specific nodes of the super Dynkin Diagrams that act like minuscule
coweights, meaning that they lead to Levi-like decompositions of these
superalgebras. Notice moreover that we focused on the fundamental
representation for all the symmetries treated here, indicating that the
superspin states of the super-atoms of the super chains are represented in
the fundamental.\newline
The graded L-operators obtained here are to our knowledge, still missing in
the superspin chain literature, except for the solutions of $sl(m|n)$ chains
that were computed using degenerate solutions of the graded Yang-Baxter
equation, see eq(2.20) in \cite{FRC}. These matrices were rederived in \cite%
{slmn}\ from the 4D Chern-Simons with $SL(m|n)$ symmetry focusing on the
distinguished Dynkin diagram and by extending features of the bosonic $sl(m)$
spin chain. In this linear symmetry, all simple nodes are associated to
minuscule coweights. Here, we gave a more general expression of these super
Lax matrices for any node beyond the distinguished Dynkin diagram of $%
sl(m|n);$ see (\ref{20}) where the bosonic and fermionic oscillators are
explicitly distinguished. Notice that the bosonic L-operators of the $sl(m)$
spin chain \cite{costello,quiver} can be recovered as a special case of the
graded distinguished solutions by simply taking $n=0$.\newline
Unlike the $A\left( m|n\right) $ superalgebra, the distinguished Dynkin
diagram of the $B\left( m|n\right) $ superalgebra\ only leads to one
Levi-like decomposition associated to the distinguished Dynkin diagram given
by \textbf{Figure \ref{B1}}. The graded Lax operator of the $osp(2m|2n)$
superspin chain was constructed for this specific\ case as presented in (\ref%
{21}). This novel result has an interesting similarity with the bosonic
minuscule Lax operator of the B-type spin chain. This bosonic operator is
calculated from the 4D CS in \cite{abcde}, and is associated to the only
minuscule node of the $so(2m+1)$ algebra\ which coincides with the node $%
\varepsilon _{1}-\varepsilon _{2}$\ on the $B_{m}$\ part of the
distinguished Dynkin diagram of $B\left( m|n\right) $.\newline
For the $C(n)$ superspin chain, we also considered the distinguished Dynkin
diagram of \textbf{Figure \ref{cn},} where we have two simple nodes for
which we can calculate the L-operator using the formula (\ref{lax}). The
first super L-operator is given in (\ref{22}); it corresponds to the
fermionic node\ $\varepsilon -\mathrm{\delta }_{1}$ and only contains
bosonic oscillators. The second super L-operator (\ref{23}) is associated to
the last bosonic node $2\mathrm{\delta }_{n-1}$ which is equivalent to the
minuscule node if we only consider the bosonic $C_{n-1}$ part of the
distinguished Dynkin diagram. The minuscule L-matrix of $sp(2n)$ given in
\cite{abcde} can be recovered from the super matrix (\ref{23}) as a special
case.\newline
Finally,\ the distinguished Dynkin diagram of Figure \ref{D1}\ of the $%
D\left( m|n\right) $\ symmetry yields three Levi-like decompositions for the
$osp(2m|2n)$\ Lie superalgebra corresponding to the three bosonic nodes $%
\varepsilon _{1}-\varepsilon _{2},\varepsilon _{m-1}-\varepsilon _{m}$ and $%
\varepsilon _{m-1}+\varepsilon _{m}.$\ The first one acts in a similar
fashion to the vectorial minuscule node of the $D_{n}$\ Lie algebra, and the
resulting super Lax operator (\ref{24}) is also a generalisation of the
bosonic vector Lax operator calculated in \cite{abcde}. The other two nodes
are of spinorial nature, they are treated collectively since they lead to
the same 3-grading and eventually to the same Lax operator given in (\ref{25}%
). Their bosonic counterparts have similar properties as studied in \cite%
{costello, abcde}.
\begin{table}[t]
  \centering
  \begin{tabular}{|c|c|c|c|c|}
\hline
{\small Superalgebra} & {\small Subalgebra }$\boldsymbol{l}_{\mathbf{\mu }}$
& {\small Nilpotent }$N_{+}$ & {\small L-operators} & {\small Equations} \\
\hline
$sl_{m|n}$ & $sl_{k|l}\oplus sl_{m-k|n-l}\oplus gl_{1}$ & $%
(k+l)^{c}(m-k+n-l) $ & $\mathcal{L}_{sl(m|n)}^{\mathbf{\mu }}$ & {\small (%
\ref{20})} \\ \hline
$osp_{2m+1|2n}$ & $osp_{2m-1|2n}\oplus gl_{1}$ & $2m+2n-1$ & $\mathcal{L}%
_{B_{m|n}}^{\mathbf{\mu }_{n+1}}$ & {\small (\ref{21})-(\ref{213})} \\ \hline
$osp_{2|2n-2}$ & $\left.
\begin{array}{c}
sp_{2n-2}\oplus gl_{1} \\
sl_{1|n-1}\oplus gl_{1}%
\end{array}%
\right. $ & $\left.
\begin{array}{c}
2\left( n-1\right) \\
\frac{n(n+1)}{2}-1%
\end{array}%
\right. $ & $\left.
\begin{array}{c}
\mathcal{L}_{C(n)}^{\mathbf{\mu }_{1}} \\
\mathcal{L}_{C(n)}^{\mathbf{\mu }_{n}}%
\end{array}%
\right. $ & $\left.
\begin{array}{c}
\text{{\small (\ref{22})}} \\
\text{{\small (\ref{23})-(\ref{230})}}%
\end{array}%
\right. $ \\ \hline
$osp_{2m|2n}$ & $\left.
\begin{array}{c}
osp_{2m-2|2n}\oplus gl_{1} \\
sl_{m|n}\oplus gl_{1}%
\end{array}%
\right. $ & $\left.
\begin{array}{c}
2\left( m+n-1\right) \\
\frac{\left( m+n\right) (m+n+1)}{2}-m%
\end{array}%
\right. $ & $\left.
\begin{array}{c}
\mathcal{L}_{D_{m|n}}^{\mathbf{\mu }_{m+n}} \\
\mathcal{L}_{D_{m|n}}^{\mathbf{\mu }_{n+1}}%
\end{array}%
\right. $ & $\left.
\begin{array}{c}
\text{{\small (\ref{25})-(\ref{250})}} \\
\text{{\small (\ref{24})}}%
\end{array}%
\right. $ \\ \hline
\end{tabular}%

  \caption{Summary of the results for each superalgebra}
  \label{tableconc}
\end{table}
\newline
As extensions of this work, one can follow the demarche presented here in
order to study other superspin chains in the framework of four-dimensional
Chern-Simons gauge theory; in particular those associated to exceptional Lie
superalgebras $F(4)$, $G(3)$ and $D(1,2;\alpha )$ whose 3-gradings are given
in \cite{van1}. Interestingly, the generalization of the L-operator
construction based on Levi-like decompositions, as well as the
Costello-Yamazaki-Yagi formula \cite{costello}, would allow to obtain
solutions for all nodes of the Dynkin diagram even if it they don't
correspond\ to Levi-like decompositions or minuscule coweights. These cases
lead to 5-gradings of Lie superalgebras, as listed in \cite{van}-\cite{van2}.
\newline
To end this conclusion, we collect in Table \ref{tableconc} the expressions of the super
oscillator realisations of Lax operators for the families $A(m-1\mid n-1)$, $%
B(m\mid n)$, $C(n)$ and $D(m\mid n)$ Lie superalgebras with 3-grading
decompositions as $g=\boldsymbol{l}_{\mathbf{\mu }}\oplus N_{+}\oplus N_{-}$
where $\boldsymbol{l}_{\mathbf{\mu }}$ is Levi-like subalgebra and $N_{\pm }$
nilpotent superalgebras.

\section{Appendix: Explicit check of solutions}

This appendix is added to the revised version of our paper in order to :
\begin{itemize}
\item $\left( i\right) $ Compare our orthosymplectic super Lax matrices computed
from the 4D CS theory, to similar solutions that appeared in \textrm{\cite%
{Fra}} shortly after the first version of this paper, and that are based on
algebraic analysis.
\item $(ii)$ Comment the quantum versions of the classical orthogonal super Lax
operators given here above. Notice that the quantum upgrading of the $%
sl(m|n) $ super Lax operators calculated in 4D CS is detailed in \cite{slmn}.
\item $(iii)$ Describe the R-matrix and the RLL equation of integrability for an
orthosymplectic superspin chain system, and refer to the explicit
verification of the matrices $\mathcal{L}_{B_{m|n}}^{\mathbf{\mu }_{n+1}},$ $%
\mathcal{L}_{C(n)}^{\mathbf{\mu }_{1}},$\ $\mathcal{L}_{C(n)}^{\mathbf{\mu }%
_{n}},$\ $\mathcal{L}_{D_{m|n}}^{\mathbf{\mu }_{m+n}}$\ and $\mathcal{L}%
_{D_{m|n}}^{\mathbf{\mu }_{n+1}}$\ as appropriate solutions.
\end{itemize}
First, notice that the authors in \cite{Fra} yield two types of
orthosymplectic Lax matrices referred to as linear and quadratic. The
quadratic Lax matrix is equivalent to super Lax matrices of vector type in
our formalism, i.e. to solutions corresponding to vector-like nodes: $%
\mathbf{\mu }_{n+1}$\ of the distinguished Dynkin diagram of $B_{m|n},$ and $%
\mathbf{\mu }_{n+1}$ of $\mathfrak{DD}\left[ D_{m|n}\right] .$ The linear
Lax matrix of \textrm{\cite{Fra}}\ corresponds to spinor-like solutions
associated to the spinorial nodes, $\mathbf{\mu }_{m+n}$\ of $D_{m|n}$\ and $%
\mathbf{\mu }_{n}$\ of $C(n).$ This correspondence is collected in
following table
\begin{equation*}
\begin{tabular}{c|c}
Solutions of \textrm{\cite{Fra}} & Solutions from the 4D CS \\ \hline \hline
Eq(5.54) \  \  \ (quadratic) & \  \  \
\begin{tabular}{l}
Eq(6.43) $\  \  \mathcal{L}_{D_{m|n}}^{\mathbf{\mu }_{n+1}}$ \\
Eq(4.26) $\  \  \  \mathcal{L}_{B_{m|n}}^{\mathbf{\mu }_{n+1}}$%
\end{tabular}
\  \  \ (vector type) \\
Eq(4.12) \  \ (linear) & \  \  \  \
\begin{tabular}{l}
Eq(5.33) $\  \  \  \mathcal{L}_{C(n)}^{\mathbf{\mu }_{n}}$ \\
Eq(6.27) $\  \  \  \mathcal{L}_{D_{m|n}}^{\mathbf{\mu }_{m+n}}$%
\end{tabular}
\  \ (spinorial type) \\ \hline \hline
\end{tabular}%
\end{equation*}%
Recall here that the matrices $\mathcal{L}_{B_{m|n}}^{\mathbf{\mu }_{n+1}}$\
and $\mathcal{L}_{D_{m|n}}^{\mathbf{\mu }_{n+1}}$ have the same structure,
and only differ in their dimensions; and the same for $\mathcal{L}_{C(n)}^{%
\mathbf{\mu }_{n}}$ and $\mathcal{L}_{D_{m|n}}^{\mathbf{\mu }_{m+n}}.$ We
will focus below on the comparison between Eq(5.54) of \textrm{\cite{Fra} }%
with $\mathcal{L}_{D_{m|n}}^{\mathbf{\mu }_{n+1}},$ and Eq(4.12) of \textrm{%
\cite{Fra}} with\ $\mathcal{L}_{D_{m|n}}^{\mathbf{\mu }_{m+n}}.$ But before
that, we recall the explicit integrability condition for the orthosymplectic
superspin chains.

$\bullet $ \emph{Orthosymplectic RLL equation}\newline
To solve the RLL relations for the orthosymplectic families, we consider the
orthosymplectic R-matrix reading for the superalgebra $osp(N|M)$ with pair $%
M $, as \textrm{\cite{Arn,arn2}}
\begin{equation}
R(z)=z(z+\mathrm{\varkappa })I_{d}+\left( z+\mathrm{\varkappa }\right) P-zQ
\label{rmatrix}
\end{equation}%
We have here $2\mathrm{\varkappa }=(N-M-2),\ I_{d}$ is the identiy operator,
$P$ is the permutation operator expressed in terms of the canonical matrix
generators $e_{xy}$ as,%
\begin{equation}
P=\tsum \limits_{x,y=1}^{M+N}(-)^{\left \vert y\right \vert }\left[
e_{xy}\otimes e_{yx}\right]
\end{equation}%
and $Q$ the operator given by%
\begin{equation}
Q=\sum_{x,y}(-1)^{\left \vert x\right \vert \left \vert y\right \vert }%
\mathrm{\theta }_{x}\mathrm{\theta }_{y}\left[ e_{\bar{y}\bar{x}}\otimes
e_{yx}\right]
\end{equation}%
The label $\bar{x}$ is defined by $\left( M+N+1\right) -x,$ and the numbers $%
\left( \mathrm{\theta }_{{\small 1}},...,\mathrm{\theta }_{M+N}\right) $
take the values $\pm 1$ obeying the conditions $\mathrm{\theta }_{{\small
\bar{x}}}=\left( -\right) ^{|x|}\mathrm{\theta }_{{\small x}}.$ By replacing
into (\ref{RLLL}), the RLL equations are equivalent to the following
commutations relations%
\begin{eqnarray}
L_{xy}(z),L_{kl}(z^{\prime }) &=&\frac{\left( -1\right) ^{\left \vert
x\right \vert \left \vert y\right \vert +\left \vert x\right \vert \left
\vert k\right \vert +\left \vert y\right \vert \left \vert k\right \vert }}{%
z-z^{\prime }}\left( L_{ky}(z^{\prime })L_{xl}(z)-L_{ky}(z)L_{xl}(z^{\prime
})\right)  \label{nbg} \\
&&+\frac{1}{z-z^{\prime }+\mathrm{\varkappa }}(\delta _{k\bar{x}%
}\sum_{p=1}^{N+M}L_{py}(z)L_{\bar{p}l}(z^{\prime })\left( -1\right) ^{\left
\vert x\right \vert +\left \vert x\right \vert \left \vert y\right \vert
+\left \vert y\right \vert \left \vert p\right \vert }\mathrm{\theta }_{%
{\small x}}\mathrm{\theta }_{{\small p}}  \notag \\
&&-\delta _{l\bar{y}}\sum_{p=1}^{N+M}L_{k\bar{p}}(z^{\prime
})L_{xp}(z)\left( -1\right) ^{\left \vert p\right \vert +\left \vert y\right
\vert +\left \vert x\right \vert \left \vert k\right \vert +\left \vert
x\right \vert \left \vert p\right \vert +\left \vert y\right \vert \left
\vert k\right \vert }\mathrm{\theta }_{{\small p}}\mathrm{\theta }_{{\small y%
}})  \notag
\end{eqnarray}

$\bullet $\emph{\ Eq(5.54) of \cite{Fra} with }$\mathcal{L}_{D_{m|n}}^{%
\mathbf{\mu }_{n+1}}$\emph{\ }\newline
Although a quick look into the quadratic Lax matrix of\textrm{\  \cite{Fra}}
reveals the similarity with the structure of $\mathcal{L}_{B_{m|n}}^{\mathbf{%
\mu }_{n+1}}$ and $\mathcal{L}_{D_{m|n}}^{\mathbf{\mu }_{n+1}},$\ we think
it is interesting to explicitly show this agreement by analysing the special
notations used in \textrm{\cite{Fra}} and linking them to ours. First,
because \textrm{\cite{Fra}} provides a computational check for the
verification of the RLL equation, and second to discuss the quantum
uplifting of these orthosymplectic solutions.\newline
For the D-type superalgebra that the authors in \textrm{\cite{Fra}} note as $%
osp(2n|2m),$ where the role of the integers $n$ and $m$ is opposite to our
notation, the graded quadratic Lax matrix (Eq(5.54)) is given by%
\begin{equation}
L\left( z\right) =\left(
\begin{array}{ccc}
\xi & z\mathbf{\bar{u}}-\frac{\zeta ^{\dagger }}{2}\mathbf{\bar{u}}^{T}M &
\frac{\zeta ^{\dagger }}{2} \\
-z\mathbf{\bar{u}}+\frac{\zeta }{2}M\mathbf{\bar{u}}^{T} & z\mathcal{I}_{%
{\small 2(n+m-1)}}-M\mathbf{\bar{u}}^{T}\mathbf{u}^{T}M & M\mathbf{\bar{u}}%
^{T} \\
\frac{\zeta }{2} & -\mathbf{u}^{T}M & 1%
\end{array}%
\right)  \label{L}
\end{equation}%
where we have set for convenience%
\begin{equation}
\begin{tabular}{lll}
$\xi $ & $=$ & $z^{2}-z\left( \mathcal{N}+\kappa -1\right) +\frac{1}{4}\zeta
^{\dagger }\zeta $ \\
$\mathcal{N}$ & $=$ & $\mathbf{\bar{u}u}$ \\
$\zeta $ & $=$ & $\left( \mathbf{u}^{T}M\mathbf{u}\right) $ \\
$\zeta ^{\dagger }$ & $=$ & $\left( \mathbf{\bar{u}}M\mathbf{\bar{u}}%
^{T}\right) $%
\end{tabular}
\label{KN}
\end{equation}%
In these relations, $M$ is a graded $2\left( n-1+m\right) \times 2\left(
n-1+m\right) $ matrix given in Eq(5.63) \textrm{\cite{Fra}}. The complex $%
\mathbf{u}$ and its complex conjugate $\mathbf{\bar{u}}$ (adjoint conjugate $%
\mathbf{u}^{\dagger }$ for quantum version) are graded super-oscillators
with $2\left( n-1+m\right) $ components distributed into $2\left( n-1\right)
$ bosonic oscillators $\left( a_{i},\bar{a}_{i}\right) ,$ and $2m$ fermionic
oscillators $\left( c_{\alpha },\bar{c}_{\alpha }\right) $. These components
were ordered in \textrm{\cite{Fra}} as follows%
\begin{equation}
\begin{tabular}{lll}
$\mathbf{u}$ & $=$ & $\left(
a_{2},...,a_{n};c_{n+1},...,c_{n+m+1};a_{n+2m+1},...,a_{2n+2m-1}\right) ^{T}$
\\
$\mathbf{\bar{u}}$ & $=$ & $\left( \bar{a}_{2},...,\bar{a}_{n};\bar{c}%
_{n+1},...,\bar{c}_{n+m+1};\bar{a}_{n+2m+1},...,\bar{a}_{2n+2m-1}\right)
^{T} $%
\end{tabular}%
\end{equation}%
To make contact with our 4D CS based solution (\ref{24}), we write the
graded vectors $\left( \mathbf{u,\bar{u}}\right) $ as%
\begin{equation}
\mathbf{u}=\left( \mathbf{b},\mathbf{\beta }\right) \quad ,\quad \mathbf{%
\bar{u}}=\left( \mathbf{c},\mathbf{\gamma }\right)
\end{equation}%
or equivalently by using contravariant $u^{\text{\textsc{a}}}$ and covariant
$\bar{u}_{\text{\textsc{a}}}$ super labels like%
\begin{equation}
\left( u^{\text{\textsc{a}}}\mathbf{,}\bar{u}_{\text{\textsc{a}}}\right)
,\qquad u^{\text{\textsc{a}}}=\left( b^{i},\beta ^{\alpha }\right) ,\qquad
\bar{u}_{\text{\textsc{a}}}=\left( c_{i},\gamma _{\alpha }\right)  \label{vw}
\end{equation}%
Notice that generally speaking, the complex variables $\bar{u}_{\text{%
\textsc{a}}}$ are not necessarily the complex conjugates of $u^{\text{%
\textsc{a}}}$ [i.e: $\bar{u}_{\text{\textsc{a}}}\neq \left( u^{\text{\textsc{%
a}}}\right) ^{\dagger }$]. Also, in (\ref{vw}), the $\left(
b^{i},c_{i}\right) $ are bosonic quantities and the $\left( \beta ^{\alpha
},\gamma _{\alpha }\right) $ are fermionic; they are ordered as%
\begin{equation}
\begin{tabular}{lllllll}
$b^{i}$ & $=$ & $\left( b^{1},...,b^{2n-2}\right) $ & $,\qquad $ & $\beta
^{\alpha }$ & $=$ & $\left( \beta ^{1},...,\beta ^{2m}\right) $ \\
$c_{i}$ & $=$ & $\left( c_{1},...,c_{2n-2}\right) $ & $,\qquad $ & $\gamma
_{\alpha }$ & $=$ & $\left( \gamma _{1},...,\gamma _{2m}\right) $%
\end{tabular}%
\end{equation}%
Using this new notation, we can identify the quadratic Lax matrix of (\ref{L}%
) derived in \textrm{\cite{Fra}, with our spinor-like matrix }$\mathcal{L}%
_{D_{m|n}}^{\mathbf{\mu }_{n+1}}$\textrm{\ given by eq(\ref{24}), by using
the following correspondence}
\begin{equation}
\begin{tabular}{c|c|c}
superoscillator & bosonic & fermionic \\ \hline \hline
(\ref{L}) & $\left.
\begin{array}{ccc}
\text{{\small classical}} & : & \left( a_{i},\bar{a}_{i}\right) \\
\text{{\small quantum}} & : & \left( a_{i},a_{i}^{\dagger }\right)%
\end{array}%
\right. $ & $\left.
\begin{array}{ccc}
\text{{\small classical}} & : & \left( c_{\alpha },\bar{c}_{\alpha }\right)
\\
\text{{\small quantum}} & : & \left( c_{\alpha },c_{\alpha }^{\dagger
}\right)%
\end{array}%
\right. $ \\
\textrm{(\ref{24})} & $%
\begin{array}{ccc}
\text{{\small classical}} & : & \left( \mathrm{b}^{i}\mathrm{,c}_{i}\right)
_{i=1}^{2n-2} \\
\text{{\small quantum}} & : & (\mathrm{\hat{b}}^{i}\mathrm{,\hat{c}}%
_{i})_{i=1}^{2n-2}%
\end{array}%
$ & $\left.
\begin{array}{ccc}
\text{{\small classical}} & : & \left( \mathrm{\beta }^{\alpha }\mathrm{%
,\gamma }_{\alpha }\right) _{\alpha =1}^{2m} \\
\text{{\small quantum}} & : & (\mathrm{\hat{\beta}}^{\alpha }\mathrm{,\hat{%
\gamma}}_{\alpha }^{\dagger })_{\alpha =1}^{2m}%
\end{array}%
\right. $ \\ \hline \hline
\end{tabular}%
\end{equation}%
\begin{equation*}
\end{equation*}%
The hatted quantities $\mathrm{\hat{b}}^{i}\mathrm{,\hat{c}}_{i}$ and $%
\mathrm{\hat{\beta}}^{\alpha }\mathrm{,\hat{\gamma}}_{\alpha }^{\dagger }$
designate quantum oscillators (non super commuting operators), their non
vanishing super commutation relations are given by
\begin{equation}
\lbrack \mathrm{\hat{b}}^{i}\mathrm{,\hat{c}}_{j}]=\delta _{j}^{i}\qquad
,\qquad \left \{ \mathrm{\hat{\beta}}^{\alpha }\mathrm{,\hat{\gamma}}%
_{\sigma }\right \} =\delta _{\sigma }^{\alpha }
\end{equation}%
From the structure of (\ref{24}), we learn that we also have%
\begin{equation}
\xi =z^{2}-z\left( \mathcal{N}+\kappa -1\right) +\frac{1}{4}\zeta ^{\dagger
}\zeta
\end{equation}%
and
\begin{equation}
\begin{tabular}{lllll}
$\mathcal{N}$ & $=$ & $\bar{u}_{\text{\textsc{a}}}u^{\text{\textsc{a}}}$ & $%
= $ & $-\left( \mathrm{c}_{i}\mathrm{b}^{i}+\mathrm{\gamma }_{\alpha }%
\mathrm{\beta }^{\alpha }\right) $ \\
$\zeta $ & $=$ & $u^{\text{\textsc{a}}}G_{\text{\textsc{ab}}}u^{\text{%
\textsc{b}}}$ & $=$ & $-\left( \mathbf{b}^{2}+\mathbf{\beta }^{2}\right) $
\\
$\zeta ^{\dagger }$ & $=$ & $\bar{u}_{\text{\textsc{a}}}G^{\text{\textsc{ab}}%
}\bar{u}_{\text{\textsc{b}}}$ & $=$ & $-\left( \mathbf{c}^{2}\mathbf{+\gamma
}^{2}\right) $%
\end{tabular}%
\end{equation}%
\begin{equation*}
\end{equation*}%
which should be compared with (\ref{KN}). The metric $G_{\text{\textsc{ab}}}$
is equal to $-\delta _{\text{\textsc{ab}}},$\ and $G^{\text{\textsc{ab}}}$\
is its inverse.\newline
Notice here that the classical quantity $\mathrm{c}_{i}\mathrm{b}^{i}+%
\mathrm{\gamma }_{\alpha }\mathrm{\beta }^{\alpha }$ reading also like the
mean value%
\begin{equation}
\mathrm{c}_{i}\mathrm{b}^{i}+\mathrm{\gamma }_{\alpha }\mathrm{\beta }%
^{\alpha }=\frac{1}{2}\left( \mathrm{c}_{i}\mathrm{b}^{i}+\mathrm{b}^{i}%
\mathrm{c}_{i}\right) +\frac{1}{2}\left( \mathrm{\gamma }_{\alpha }\mathrm{%
\beta }^{\alpha }-\mathrm{\beta }^{\alpha }\mathrm{\gamma }_{\alpha }\right)
\end{equation}%
due to graded commutativity, gets mapped at the quantum level into the sum
of the operators $(\mathrm{\hat{c}}_{i}\mathrm{\hat{b}}^{i}+\mathrm{\hat{b}}%
^{i}\mathrm{\hat{c}}_{i})/2+$ $(\mathrm{\hat{\gamma}}_{\alpha }\mathrm{\hat{%
\beta}}^{\alpha }+\mathrm{\hat{\beta}}^{\alpha }\mathrm{\hat{\gamma}}%
_{\alpha })/2$. By using the graded commutation relations, one generates a
vaccum contribution as shown below
\begin{equation}
\mathrm{\hat{c}}_{i}\mathrm{\hat{b}}^{i}+\mathrm{\hat{\gamma}}_{\alpha }%
\mathrm{\hat{\beta}}^{\alpha }+\frac{2\left( n-1\right) -2m}{2}=\mathrm{\hat{%
c}}_{i}\mathrm{\hat{b}}^{i}+\mathrm{\hat{\gamma}}_{\alpha }\mathrm{\hat{\beta%
}}^{\alpha }+\frac{2\left( n-m\right) }{2}-1
\end{equation}%
This expression indicates that the number $\kappa $ in the super Lax matrix (%
\ref{L}) is just the quantity $2n-2m$ which sometimes is termed as the
Maxwell-Callading index.

$\bullet $\emph{\ Eq(4.12) of \cite{Fra} with }$\mathcal{L}_{D_{m|n}}^{%
\mathbf{\mu }_{n+m}}$\emph{\ }\newline
Regarding the identification of the orthosymplectic rational RLL solution
given in matrix form in eq(4.12) of \textrm{\cite{Fra},} with the
spinor-like super Lax matrix $\mathcal{L}_{D_{m|n}}^{\mathbf{\mu }_{n+m}},$
notice that they are equivalent by the multiplication of $\mathcal{L}%
_{D_{m|n}}^{\mathbf{\mu }_{n+m}}$\ by the overall factor $z^{-1},$\ which is
permitted by the RLL equation symmetries. The two matrices are represented
in similar bases dividing the $2m+2n$ dimensions as $(m+n,\overline{m+n}),$
the difference is that the integers $n$ and $m$ play opposite roles in \textrm{\cite{Fra}} where the superalgebra is taken as $%
D(n|m)=osp(2n|2m)$\ (instead of $osp(2m|2n)$\ for us). However, since our
basis is ordered as $\left( m,n,\overline{n},\overline{m}\right) ,$\ and
theirs as $\left( n,m,\overline{m},\overline{n}\right) ,$\ we have lookalike
matrices where we have the correspondence\
\begin{equation}
\begin{tabular}{|c|c|}
\hline
$\mathcal{L}_{D_{m|n}}^{\mathbf{\mu }_{n+m}}$ from 4D CS Eq(\ref{25}) &
Eq(4.12) of \textrm{\cite{Fra}} \\ \hline
$\Phi $ & $\overline{\mathbf{A}}$ \\ \hline
$\Lambda $ & $\overline{\mathbf{B}}$ \\ \hline
$\Psi $ & $-\mathbf{A}$ \\ \hline
$\Gamma $ & $\mathbf{B}$ \\ \hline
$\mathrm{\gamma }_{\bar{\alpha}i}$ & $\mathbf{-C}$ \\ \hline
$\mathrm{\beta }^{i\bar{\alpha}}$ & $\overline{\mathbf{C}}$ \\ \hline
\end{tabular}%
\end{equation}%
Finally, notice that the quadratic graded matrix (5.54) of \textrm{\cite{Fra}%
}\ is proven as a solution of the RLL equation with the orthosymplectic
R-matrix (\ref{rmatrix}); which further verifies the matrices $\mathcal{L}%
_{D_{m|n}}^{\mathbf{\mu }_{n+1}}$ and $\mathcal{L}_{B_{m|n}}^{\mathbf{\mu }%
_{n+1}}$ as super Lax matrices for the orthosymplectic superspin chains of $B
$ and $D$ type. Moreover, the elements of the matrix (4.12) of \textrm{\cite%
{Fra}}\ verify the commutation relations (\ref{nbg}), and thus\ provide a
check for the solutions $\mathcal{L}_{D_{m|n}}^{\mathbf{\mu }_{m+n}}$ and $%
\mathcal{L}_{C(n)}^{\mathbf{\mu }_{n}}$ of our paper. The matrix $\mathcal{L}%
_{C(n)}^{\mathbf{\mu }_{1}}$ is a special solution because it is purely
bosonic. It should be verified by using the non-graded version of the
orthosymplectic R-matrix, which is given by the orthogonal R-matrix Eq(2.1)
in \cite{FRASSEK2}. Indeed, this matrix is equal to the verified Lax matrix
of the orthogonal $so(2n)$ spin chain given in eq(3.42)\ of \cite{abcde};
see also eq(4.12) of \cite{FRASSEK2}.%
\begin{equation*}
\end{equation*}

\end{document}